\documentclass[twocolumn]{aastex62}

\graphicspath{{./}}

\received{January 1, 2018}
\revised{January 7, 2018}
\accepted{\today}
\submitjournal{ApJ}

\shorttitle{Extended Aperture Photometry of K2 RR Lyrae stars}
\shortauthors{Plachy et al.}

\begin{document}

\title{Extended Aperture Photometry of K2 RR Lyrae stars}

\correspondingauthor{E. Plachy and L. Moln\'ar}
\email{eplachy@konkoly.hu}
\email{lmolnar@konkoly.hu}

\author[0000-0002-5481-3352]{Emese Plachy}
\affiliation{Konkoly Observatory, Research Centre for Astronomy and Earth Sciences, Konkoly Thege Mikl\'os \'ut 15-17, H-1121 Budapest, Hungary}
\affiliation{MTA CSFK Lend\"ulet Near-Field Cosmology Research Group, Konkoly Thege Mikl\'os \'ut 15-17, H-1121 Budapest, Hungary}

\author[0000-0002-8159-1599]{L\'aszl\'o Moln\'ar}
\affiliation{Konkoly Observatory, Research Centre for Astronomy and Earth Sciences, Konkoly Thege Mikl\'os \'ut 15-17, H-1121 Budapest, Hungary}
\affiliation{MTA CSFK Lend\"ulet Near-Field Cosmology Research Group, Konkoly Thege Mikl\'os \'ut 15-17, H-1121 Budapest, Hungary}

\author[0000-0002-8585-4544]{Attila B\'odi}
\affiliation{Konkoly Observatory, Research Centre for Astronomy and Earth Sciences, Konkoly Thege Mikl\'os \'ut 15-17, H-1121 Budapest, Hungary}
\affiliation{MTA CSFK Lend\"ulet Near-Field Cosmology Research Group, Konkoly Thege Mikl\'os \'ut 15-17, H-1121 Budapest, Hungary}

\author[0000-0002-7602-0046]{Marek Skarka}
\affiliation{Department of Theoretical Physics and Astrophysics, Masaryk University, Kotl\'{a}\v{r}sk\'{a} 2, 61137 Brno, Czech Republic}
\affiliation{Astronomical Institute, Czech Academy of Sciences, Fri\v{c}ova 298, 25165, Ond\v{r}ejov, Czech Republic}
\affiliation{Konkoly Observatory, Research Centre for Astronomy and Earth Sciences, Konkoly Thege Mikl\'os \'ut 15-17, H-1121 Budapest, Hungary}

\author[0000-0002-5781-1926]{P\'al Szab\'o}
\affiliation{Konkoly Observatory, Research Centre for Astronomy and Earth Sciences, Konkoly Thege Mikl\'os \'ut 15-17, H-1121 Budapest, Hungary}
\affiliation{MTA CSFK Lend\"ulet Near-Field Cosmology Research Group, Konkoly Thege Mikl\'os \'ut 15-17, H-1121 Budapest, Hungary}

\author[0000-0002-3258-1909]{R\'obert Szab\'o}
\affiliation{Konkoly Observatory, Research Centre for Astronomy and Earth Sciences, Konkoly Thege Mikl\'os \'ut 15-17, H-1121 Budapest, Hungary}
\affiliation{MTA CSFK Lend\"ulet Near-Field Cosmology Research Group, Konkoly Thege Mikl\'os \'ut 15-17, H-1121 Budapest, Hungary}

\author[0000-0003-0700-6176]{P\'eter Klagyivik }
\affiliation{Instituto de Astrof\'isica de Canarias, C. V\'ia L\'actea S/N, 38205 La Laguna, Tenerife, Spain}
\affiliation{Universidad de La Laguna, Departamento de Astrof\'isica, Av. Astrofisico Francisco S\'anchez, S/N, 38206 San Crist\'obal de La Laguna, Tenerife, Spain}
\affiliation{Konkoly Observatory, Research Centre for Astronomy and Earth Sciences, Konkoly Thege Mikl\'os \'ut 15-17, H-1121 Budapest, Hungary}

\author{\'Ad\'am S\'odor}
\affiliation{Konkoly Observatory, Research Centre for Astronomy and Earth Sciences, Konkoly Thege Mikl\'os \'ut 15-17, H-1121 Budapest, Hungary}
\affiliation{MTA CSFK Lend\"ulet Near-Field Cosmology Research Group, Konkoly Thege Mikl\'os \'ut 15-17, H-1121 Budapest, Hungary}

\author[0000-0003-2595-9114]{Benjamin J. S. Pope}
\affiliation{Center for Cosmology and Particle Physics, Department of Physics, New York University, 726 Broadway, New York, NY 10003, USA}
\affiliation{Center for Data Science, New York University, 60 Fifth Ave, New York, NY 10011, USA}
\affiliation{NASA Sagan Fellow}



\begin{abstract}

The \textit{Kepler} space telescope observed thousands of RR Lyrae stars in the K2 mission. In this paper we present our photometric solutions using extended apertures in order to conserve the flux of the stars to the highest possible extent. With this method we are able to avoid most of the problems that RR Lyrae light curves produced by other pipelines suffer from. For post-processing we apply the K2SC pipeline to our light curves. We provide the EAP (Extended Aperture Photometry) of 432 RR Lyrae stars observed in campaigns 3, 4, 5, and 6. We also provide subclass classifications based on Fourier parameters. We investigated in particular the presence of the Blazhko effect in the stars, and found it to be 44.7\% among the RRab stars, in agreement with results from independent samples. We found that the amplitude and phase modulation in the Blazhko stars may behave rather differently, at least over the length of a K2 Campaign. We also identified four anomalous Cepheid candidates in the sample one of which is potentially the first Blazhko-modulated member of its class. 

\end{abstract}

\keywords{RR Lyrae variable stars (1410), Light curves (918), Space telescopes (1547)}


\section{Introduction} 
RR Lyrae stars are old, population II, core helium-burning stars. They occupy the intersection of the horizontal branch and the classical instability strip in the Hertzsprung-Russell diagram. They pulsate predominantly in a few radial modes (fundamental, first-overtone, or both). RR Lyrae stars are excellent tracers of old galactic populations: their large-amplitude, characteristic variations are easy to identify, and their luminosities are very similar in the visual band, making them excellent distance indicators. 

RR Lyrae stars were considered ``simple" pulsators for a long time, limited to radial modes that were only complicated by the occasional appearance of the Blazhko effect, a (quasi-)periodic modulation of the pulsation amplitude and phase, discovered over a century ago \citep{blazhko}. This picture, however, changed dramatically since the 2000s, with the start of more extensive ground-based surveys and the rise of photometric space telescopes. The first discoveries, e.g., the prevalence of the Blazhko effect, multiperiodic modulations, the detection of low-amplitude additional modes, came from various programs, such as the Konkoly Blazhko Survey, the Optical Gravitational Lensing Experiment (OGLE) and the \textit{MOST} and \textit{CoRoT} missions \citep{most,kbs,corot,ogleiii,ogleiv}. But the biggest driver of the revolution in our knowledge about RR Lyrae stars has been the \textit{Kepler} mission. The four-year-long observations provided the clearest views of the new phenomena, and inspired new theoretical developments \citep[see, e.g.,][]{bk11,Benko-2014,Moskalik-2015,dziem-2016}. 

However, the field-of-view of the original \textit{Kepler} mission was limited to about 40 RR~Lyrae stars, and did not even include all subtypes of RR Lyrae variables. In the second mission of the \textit{Kepler} space telescope, astronomers have had a golden opportunity to propose many more targets for observation along the Ecliptic Plane. The great cooperation between the K2 Guest Observer Office and the astronomical community has lead to an unprecedented photometric data set of various stellar objects that will be hard to exceed in quantity, quality or even the length of continuous observations. The members of Working Group 7 have proposed thousands of RR Lyrae stars and hundreds of Cepheids (for details of target selection method and the success rates of proposals we refer to \citet{Plachy_CoKon}). Results of the boutique analysis of some targets have been already published \citep{Kurtz-2015,Molnar-2015a,plachy_rrd,bodi2018} as well as of the early K2-E2 engineering data of RR Lyrae stars \citep{Molnar-2015b}. Nevertheless extensive study or population analysis based on K2 RR Lyrae data is not available yet, and the reason of this is the lack of precise light curves that minimize instrumental effects while preserving the variation of the stars. The validation work of the targets common between \textit{Kepler/K2} observations and the \textit{Gaia} DR2 variable candidates only focussed on the identification of these variables \citep{k2gaia}. Recently, a rather controversial claim was made by \citet{kovacs2018} that nearly all stars observed in the \textit{K2} mission show the Blazhko effect (or, at least, side peaks in the Fourier spectra). This study, based on the light curves released by the mission, clearly at odds with prior studies that put the occurrence rate at about 50\% \citep{kbs,Benko-2014}. 

Clearly, these stars hold questions that need the best light curves achievable to answer. Our aim is to build up a photometric database from the K2 observations that may serve as a golden sample of high-precision RR Lyrae light curves, compounding the few longer light curves from the prime mission \citep{Benko-2014}. We have chosen the RR Lyrae candidates from campaign 3 to 6 from Guest Observer proposals GO3040, GO4069, GO5069 and GO6082. These campaigns do not suffer from instrumental issues  characterize the early campaigns, the off-nominal pointing at the beginning and the mid-campaign breaks. 
In section 2 we present the problems of automatic photometry that makes them suboptimal for RR Lyrae analysis. In section 3 we present our method with step-by-step description. In section 4 we discuss the quality of our data comparing with other existing ones. 


\section{Problems to solve}
\label{sect2}
We note that RR Lyrae light curves can be notoriously hard to reduce and correct, and some of the pipelines listed below were developed as general-purpose methods or with emphasis towards planetary transits and solar-like oscillations. Therefore our assessments are indicative only of the RR Lyrae light curves they produce, and not of the overall quality of these data sets. 

A quick summary of the challenges RR Lyrae light curves may pose:

\begin{itemize}
\item Large amplitudes: the photometric amplitude of the main mode reaches 0.5-1.0 mag. However, if the radial mode is filtered out, additional modes can be detected even in the sub-mmag range. Variation has to be preserved across about 5 magnitudes of flux changes, down to the sub-mmag level.
\item Pulsation periods of RR Lyrae stars range from about 6 to 24 hr, clustering around 12-14 hr. This means that for many stars, the pulsation periods are uncomfortably close to the (or twice the) period of the K2 attitude control manoeuvres, making their separation difficult. 
\item Sharp features: light curves of fundamental-mode stars can be pronouncedly sawtooth-shaped, with a steep rising branch, and sharp maximum, followed by a slow ascending branch. Generic methods can be ill-equipped to follow these variations.
\end{itemize}

As we can see, the large pulsation signal in the RR~Lyrae light curves can be hard to remove if the data is oversmoothed. Yet it is necessary as it may overwhelm any instrumental signals that nevertheless impact the low-amplitude components of the frequency spectrum. 

Now let us look at the instrumental effects of the K2 data itself that needs to be corrected. The common issue for all K2 data comes from the two reaction-wheel mode of the space telescope. Without the third reaction wheel, the telescope rotated freely about its optical axis due to the torque caused by the uneven distribution of solar radiation pressure compared to its center of mass. The telescope was set to the unstable equilibrium point to minimize the torque, but attitude control manoeuvres with thrusters were still required after about 6 (sometimes 12) hours to return it to the desired attitude. The direction of the roll and reset movements were reversed during the campaign, usually (but not always) once, around the middle of the campaign. These movements introduced drifts and sudden jumps in the positions of the stellar images over the CCDs that got stronger towards the edges of the field-of-view. Field-of-views along the Ecliptic could be observed for up to 80 days before the telescope had to reorient itself to keep the solar panels pointed towards the Sun. 

The rotation of the field-of-view introduced two types of systematic errors: first, nearby stars can contaminate each other's apertures (or, alternatively, flux can be lost from too tight apertures). Second, even small movements introduce flux level changes caused by intra- and interpixel sensitivity variations. Sensitivity varies slightly both from pixel to pixel and from the center to the edges of any individual pixel. Although \textit{Kepler} has an initial, large-scale flat-field model, created before the launch of the spacecraft, it has no means of obtaining detailed flat-field images in space. The original mission was able to largely avoid the issue of pixel-level sensitivity variations by providing very precise, sub-pixel pointing stability. In the K2 mission, however, variations in pixel sensitivity levels created clear systematic, sawtooth-shaped flux variations, as the images of stars moved about. Therefore several techniques have been developed to correct systematic errors, and we briefly summarize their efforts and limitations below. 

\paragraph{Baseline SAP and PDCSAP photometry} The mission provides Simple Aperture Photometry (SAP) and Pre-search Data Conditioned SAP (PDCSAP) light curves, extracted from the Target Pixel Files (TPF) downloaded from the spacecraft \citep{pdcsap}. The former is simply the photometry of selected target pixels, and thus includes the systematic errors described above. Apertures are calculated for each individual frame and then an optimal aperture is created, balancing between coverage and contamination from the background (e.g. ignoring pixels that are used by very few frames). The PDCSAP photometry creates basis vectors from selected regions of the field-of-view that describe the components of the most common photometric variations across multiple channels. These are then fitted and removed from the data. While this correction works quite well, the assigned apertures can be too tight, especially towards the ends of the campaigns where the roll intensifies, and strong flux loss may occur. Although the PDCSAP algorithm is able to compensate for this, the results are less than optimal.

\paragraph{K2SFF Extracted Lightcurves} This was the first correction method published \citep{k2sff}. The K2SFF pipeline applies a self-flatfielding (SFF) algorithm to the data, exploiting the fact that stars move back and forth over the same paths multiple times, and thus the pixel sensitivity variations can be mapped from the flux variations. The method assigns circular apertures to the faint objects and more well-defined, PRF-based apertures to brigher ones (Pixel Response Function, see \citealt{bryson-2010}). This method filters out slow variations from the light curves before decorrelating the flux variations from the variations in centroid positions. However, the variations of RR Lyrae stars are faster than their 1.5-day time steps for spline fitting, and the flux changes caused by stellar pulsation far outweigh the instrumental variations. Therefore the K2SFF algorithm is, in almost all cases, incapable to correct the light curves, and mostly just introduces bogus variations into them instead. 

\paragraph{K2 Variability Catalog (K2VARCAT)} This method uses a very similar self-flatfielding (or decorrelating) algorithm as the previous one, but it also includes a machine-learning algorithm to classify the stars into variability types \citep{k2varcat,k2varcat-vars}. It is available for C0-C4 at the time of writing, and uses circular apertures that are centered to the brightest pixel, and their radii are based on the brightness of the stars. The background subtraction was performed with the median value of pixels outside the aperture, and stellar variability is removed with boxcar-fitted, 3rd degree polynomials. The detrending is based on decorrelating the variation of the centroid position and the raw flux. This method works well in most of the cases, if stellar variability occurs on different time scale than the drift and the intrinsic variability do not dominate over the pointing noise. Unfortunately, for RR Lyrae stars, this method also does more harm than good.

\paragraph{K2 Pixel Photometry (K2P$^2$)} This photometric pipeline was created by the Kepler Asteroseismic Science Operations Center \citep[KASOC,][]{k2p2}. This method uses a more detailed procedure to identify stars in the TPFs and assign apertures to them. After initial detection, blended stellar images are separated with the watershed method, and apertures are created for each one. The watershed algorithm finds the borders between different stars in the pixel mask by using the idea that the image is a topological surface. It inverts the image and considers the stars as neighbouring basins, then starts to fill them up with water from the bottom. Where the water levels of the basins meet, the watershed will be the separator line between the two stars \citep{water}.
This way the K2P$^2$ pipeline generates photometry not only for the target star in the TPF, but also for any other background objects. This was especially important for the early campaigns (C0, C1, C2) where TPFs used large masks to accommodate potentially larger deviations in the pointing \citep{k2p2-2}. Stellar variation is identified with the so-called KASOC filter that was optimized for solar-like oscillations, and then the instrumental variations are corrected with a flux-pixel position decorrelation method similar to the ones described above. 

\paragraph{K2 Systematics Correction (K2SC)} Unlike the previous pipelines, K2SC does not start from the TPFs: it is rather a post-processing algorithm for the photometry extracted from the TPFs \citep{k2sc2,k2sc}. Their published light curves are based on the SAP and PDCSAP files, but custom-mask photometry made with PyKE can also be processed. This pipeline uses the non-parametric Gaussian-process regression method to simultaneously fit both the time-dependent variation (the assumed astrophysical signal) and the position-dependent instrumental effects. A drawback of the otherwise powerful method is its reliance on the SAP/PDCSAP light curves, however, it can be used very efficiently with custom-aperture light curves, as we will show it later.

\paragraph{EPIC Variability Extraction and Removal for Exoplanet Science Targets (EVEREST)} This pipeline uses yet another method to correct for the attitude changes. In this case, instead of assigning apertures to the stars and making assumptions about the expected stellar and instrumental variations, a pixel-level decorrelation (PLD) is employed \citep{everest}. The basic principle of PLD is that astrophysical signals are the same in all the pixels corresponding to one star, whereas instrumental effects, such as pointing drifts, are not. Stellar variability is fitted with Gaussian processes, and the PLD is applied to the residual. Here the main drawback is that the Gaussian process may capture the instrumental variations alongside with the pulsation of the RR Lyrae stars, given their similar timescales, and thus they are insufficiently removed or further artifacts arise. The current (2.0) version of the pipeline overfits close to half of the RR Lyrae light curves \citep{everest2}. 

\paragraph{Planet candidates from OptimaL Aperture Reduction} Later, another group published a new set of light curves called POLAR) \citep{polar}. This work was dedicated to identify transit-like features, e.g., it was not a general-purpose pipeline. It used the aperture selection algorithm originally developed for the imagettes in the CoRoT mission, and and SFF algorithm similar to the one created by \citet{k2sff}. A visual comparison of the light curves indicates that for RR Lyrae stars, this pipeline suffers from the same obvious problems as the other SFF-based photometries, therefore we decided that further tests are not necessary. 

\section{The EAP method}
The main difference between the approaches listed above and our own method is that we assign the pixel masks individually to the stars, by hand. The origins of this method go back to the analysis of the K2-E2 observations \citep{Molnar-2015b}. The most critical and time-consuming part of this method is to find the ideal target pixel masks, but since the RR~Lyrae sample in most campaigns are limited to 1-200 stars, it is manageable. Our tests show that a careful selection of pixels can highly improve the quality of the data, especially for faint and/or contaminated stars, and it can be crucial in the investigations of small amplitude additional modes as we demonstrate it in Section 4. 

We note that towards the end of this work, a new software tool has been developed by the Kepler Guest Observer Office. The \textsc{lightkurve} package now offers interactive tools to render light curves based on the selected pixels on the fly, making future applications of our algorithm much faster and easier \citep{lightkurve}. 

Based on that, in a forthcoming paper (Szab\'o et al. in prep.) we will present an automated procedure for the rest of the K2 Campaigns, where the RR Lyrae sample published in this work is used as a training sample for our automated method.

\subsection{Creating the pixel masks}

Our main principle in creating the pixel masks is to contain the movements of the Point Spread Function (PSF), i.e., the image of the star, caused by the drift of the telescope, and thus conserve the stellar flux throughout the campaign. In most cases this can be done with some expense of slightly increased background flux. We note, however, that without invoking more complicated methods (e.g., weighted apertures based on CCD positions), there will always be a trade-off between the slightly increased background noise from a more extended aperture, and some flux (and therefore information) loss caused by a too tight aperture that leads to lower photometric accuracy.

\begin{figure}
\begin{center}
\resizebox{42mm}{!}{\includegraphics{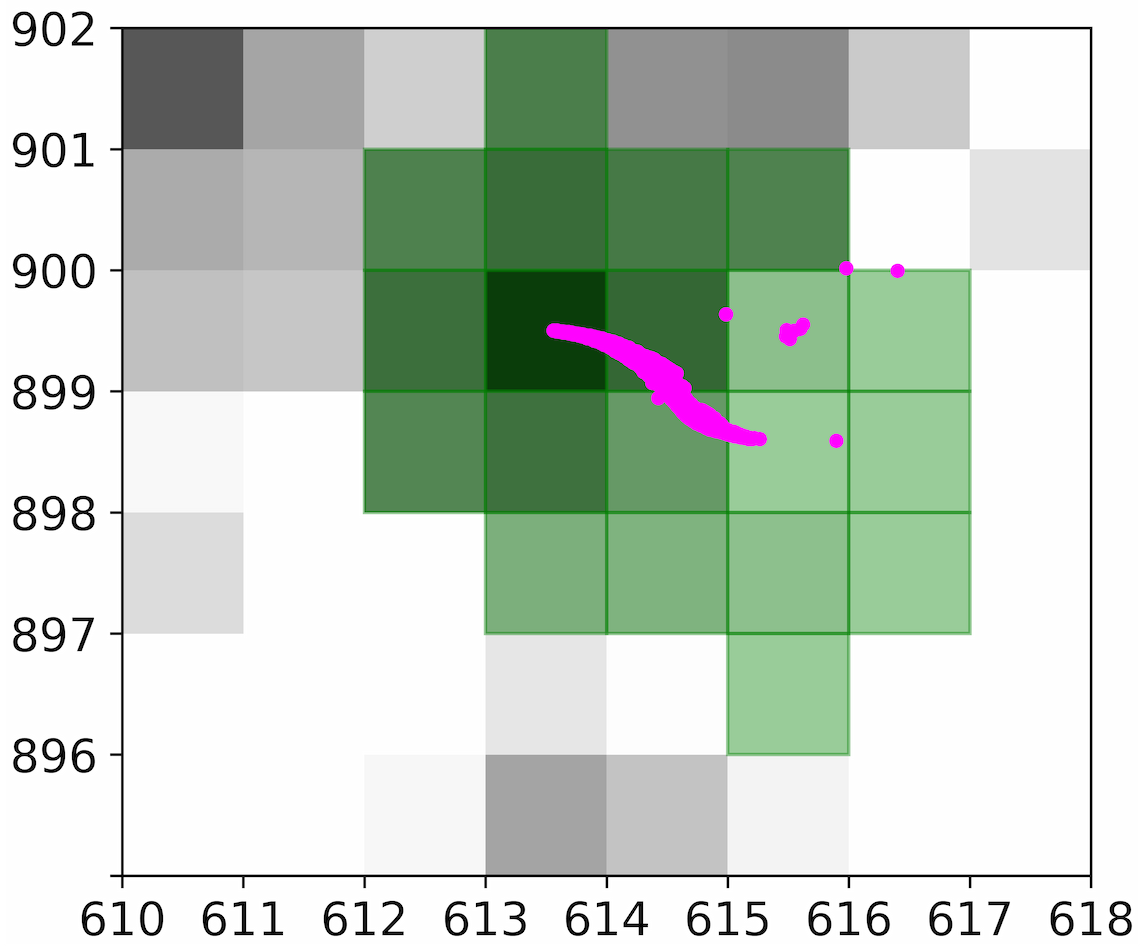}}\hspace*{0mm}%
\resizebox{42mm}{!}{\includegraphics{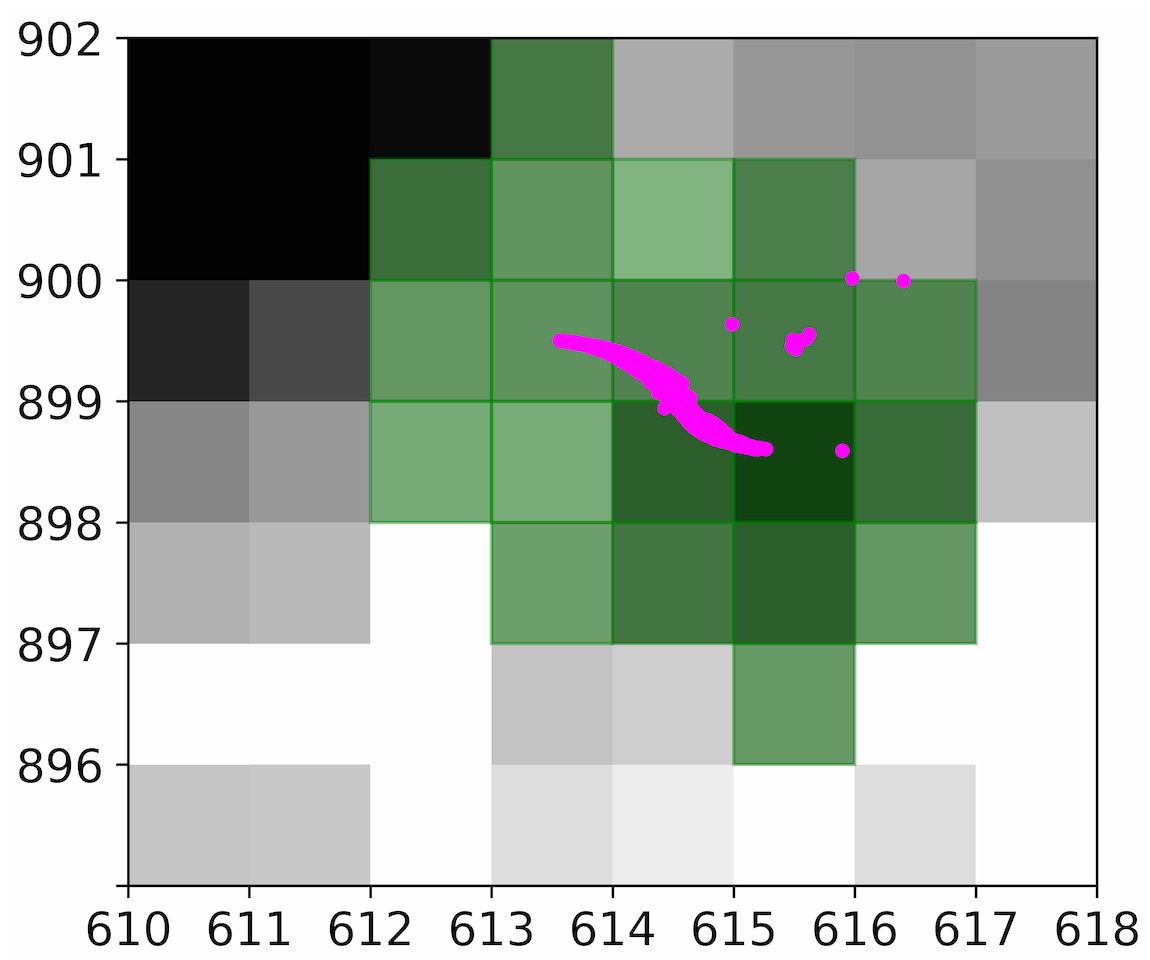}}\hspace*{0mm}%
\end{center}
\caption{The EAP mask (green) and photocenter positions (pink) of EPIC 212789806. The two images were taken at the extreme positions. Axes denote pixel row and column numbers.}
\label{fig:mask}
\end{figure}

We used the {\sc pyke} software to manually create the pixel masks \citep{pyke}. We used the $x$ and $y$ coordinates of the centroid positions to check if the mask covers all the pixels that contain the stellar flux in the extrema of the positions (Fig.~\ref{fig:mask}). The largest deviations in positions occur at the same time within a certain campaign for all stars, but we also had to take the pulsation of the star into account. Since RR~Lyrae stars have large amplitudes, the sizes of the PSFs can change considerably over a pulsation cycle, especially for brighter stars. In Figure 1 we display examples of masks that we used in EAP. 

It is, however, not always straightforward to determine the extent of the pixel masks. Nearby stars may blend together, or their movements may overlap on the CCD, contaminating each other's pixel masks. Instrumental signals, such as frame transfer and video crosstalk patterns also contaminate the images of a few stars \citep{handbook}. If blends cannot be separated, we decided to include both stars in the pixel mask. 
If the stars can be separated reasonably, e.g., contamination from other stars only affect the extreme positions, and/or the edge of the PSFs, we excluded those pixels. While the blended solutions contain the variation of the contaminant star, these are almost always orders of magnitude smaller than that of the main mode(s) of RR~Lyrae pulsations and may appear at different frequency ranges. However, low-amplitude additional modes can be still contaminated. 

\subsection{Correction of the photometry with K2SC}

Although our EAP apertures were sufficient to capture the stellar flux throughout the campaigns for a majority of the stars, the light curves were not free of instrumental effects, as two important sources of systematical variations still occurred. One was caused by the pixel sensitivity variations, as mentioned above. The other one is the use of discrete, integer pixels in {\sc pyke} to define the aperture instead of circular or elliptical apertures that would be always centered at the photocenter of the star, and would sample the PSF more evenly. This latter effect is much smaller than the former, and the two were caused by the same motions and hence the are correlated.  Furthermore, a small number of stars was also affected by reflections and crosstalk patterns in the images. 

These instrumental effects manifest in the Fourier spectra as both discrete peaks and elevated noise levels. The discrete peaks appear at the integer multiplets of the attitude correction frequency of \textit{Kepler} which is $f_{\rm corr}\approx 4.06$ d$^{-1}$, and at the lowest frequency peak near zero, representing a slow drift of the images throughout the campaign. The diffuse component consists of red noise in the low-amplitude range, dominating below 4 d$^{-1}$, that drops below the white noise towards higher frequencies. While intrinsic (low-amplitude) signals that coincide with $n\,f_{\rm corr}$ are likely inseparable from the instrumental ones, the red noise component can hide additional pulsation and/or modulation peaks that can be recovered after correcting for the signals induced by the attitude changes.

As we have shown in Sect.~\ref{sect2}, there are various methods to correct for these systematic variations. We did some experiments with the SFF task implemented in \textsc{PyKE}, but the spline fit in it could not separate the variations of the RR~Lyrae stars from the instrumental ones and thus we could no produce flattened light curves for the SFF algorithm to fit. 

We instead selected the K2SC algorithm developed by \citet{k2sc}. As we briefly explained earlier, K2SC uses Gaussian processes to separate the intrinsic, stellar variations and the instrumental ones against the pixel position data of the individual stars. Although the official K2SC data release relies on the SAP/PDCSAP photometry, the algorithm could be applied to any custom mask photometry obtained with \textsc{PyKE} from Campaign 3 upwards. Therefore we run K2SC on all of our EAP light curves from Campaign 3 to 6.

\begin{figure*}
\includegraphics[width=1\textwidth]{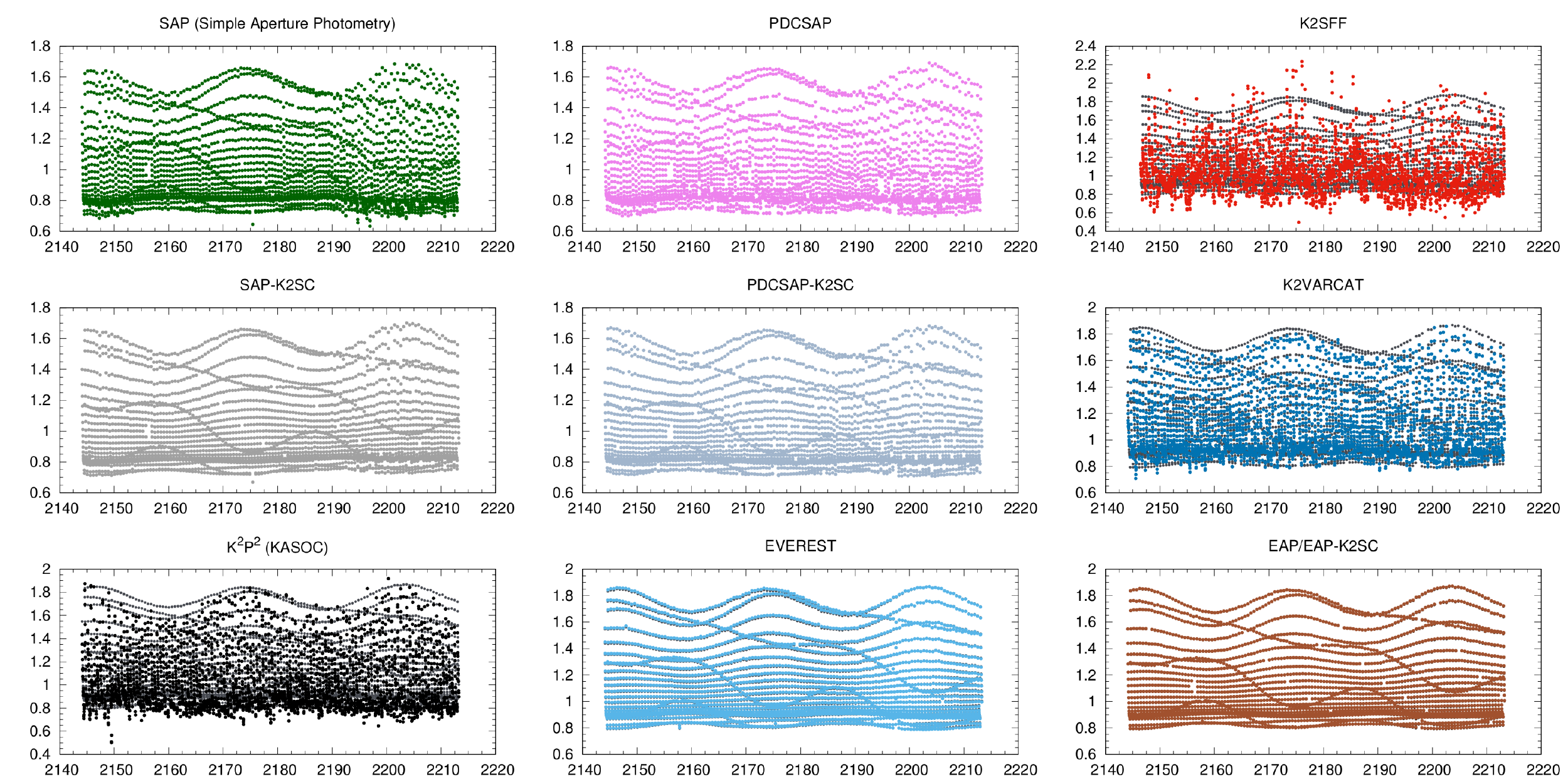}
\caption{Photometry of a relatively bright, Blazhko-modulated RRab star (EPIC 206144794, \textit{Kp} = 12.667 mag) by various pipelines. For the K2SFF, K2VARCAT, K2P$^2$, EVEREST, and EAP light curves, grey points behind the colored ones show the raw aperture photometry before the respective corrections. Individual pulsation cycles are too short to be discernible, the wavy lines are Moir\'e patterns caused by the interplay between the sampling rate and the pulsation period, the slow change in amplitude is the Blazhko effect itself. 
(x-axis: BJD-2454833, y-axis: relative flux)  }
\label{fig:k2-example1}
\end{figure*}

\begin{figure*}
\includegraphics[width=1\textwidth]{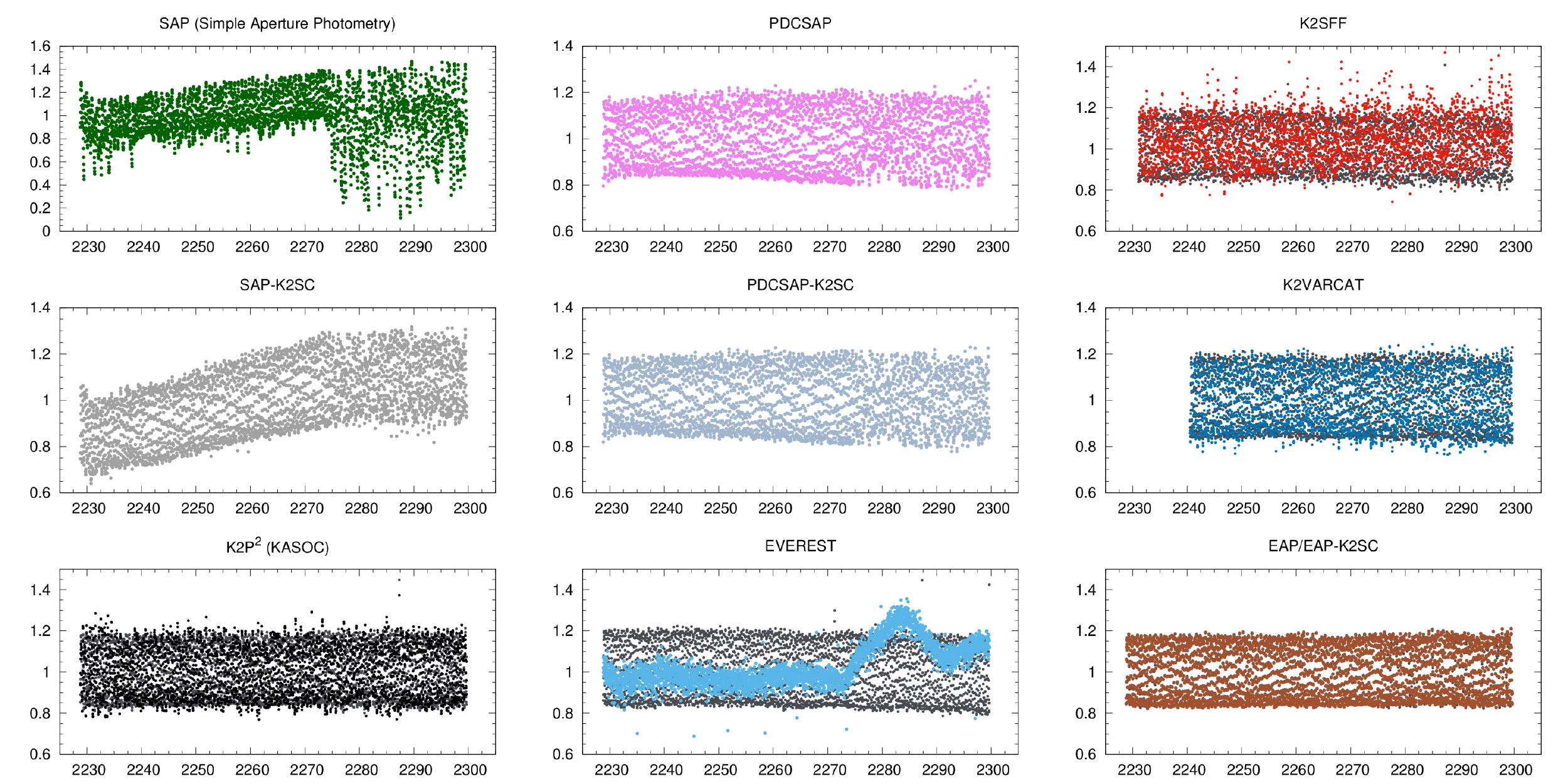}
\caption{Photometry of a relatively faint RRc star (EPIC 210438688, \textit{Kp} = 17.387 mag) by various pipelines. The too small aperture caused severe flux loss in the SAP data: this then propagated into the PDCSAP and K2SC light curves. (x-axis: BJD-2454833, y-axis: relative flux)}
\label{fig:k2-example2}
\end{figure*}

\subsection{Outlier detection, systematics}
After we applied the K2SC correction, we identified and removed outliers with simple sigma clipping. We used 3 or 4$\sigma$ limits on the residual light curves for faint and bright targets. A higher limit was deemed necessary for the brighter targets in order to preserve the points close to maximum light. For the faintest targets we had to apply iterative 3$\sigma$ clipping.

We then inspected the light curves visually to identify any remaining slow instrumental variations. These could originate from a variety of causes, either from intrinsic variations or instrumental effects, including contamination from nearby stars or reflections, crosstalk in the electronics, or poor background determination. We determined cubic spline curves for stars clearly affected and removed these signals from the K2SC-corrected light curves.

However, we did not correct the light curves for slow variations automatically, in order to preserve intrinsic signals. The limited span of the observations mean that for many stars, the modulation periods are comparable or much longer than the length of the data. It is well established that the average brightness of Blazhko RR Lyrae stars vary with the modulation period \citep{alcock2003,jurcsik2005}. These slow variations are often hard to separate from any small instrumental trends, so simple fits to low-pass filtered light curves would remove them both. Therefore we did not apply spline smoothing for any of the Blazhko stars in our sample.

\subsection{Comparison of individual light curves}

\begin{figure*}
\includegraphics[width=1\textwidth]{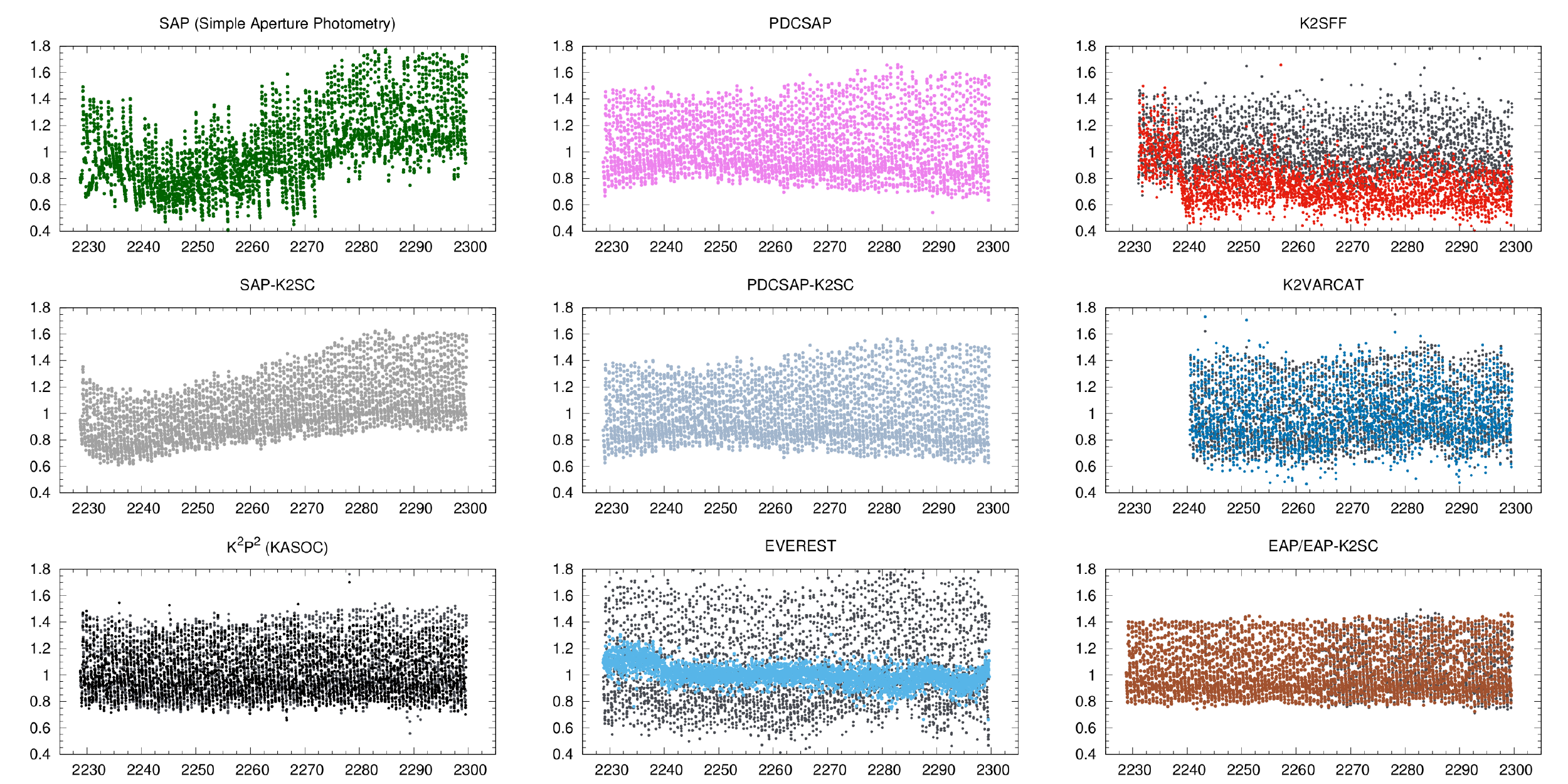}
\caption{Photometry of a faint RRab star (EPIC 210507908, \textit{Kp} = 19.085 mag) from various pipelines. Note the apparent amplitude change in the SAP, PDCSAP and K2SC light curves which may lead to a false positive detection of the Blazhko effect. (x-axis: BJD-2454833, y-axis: relative flux)} 
\label{fig:k2-example3}
\end{figure*}

\begin{figure*}
\includegraphics[width=1\textwidth]{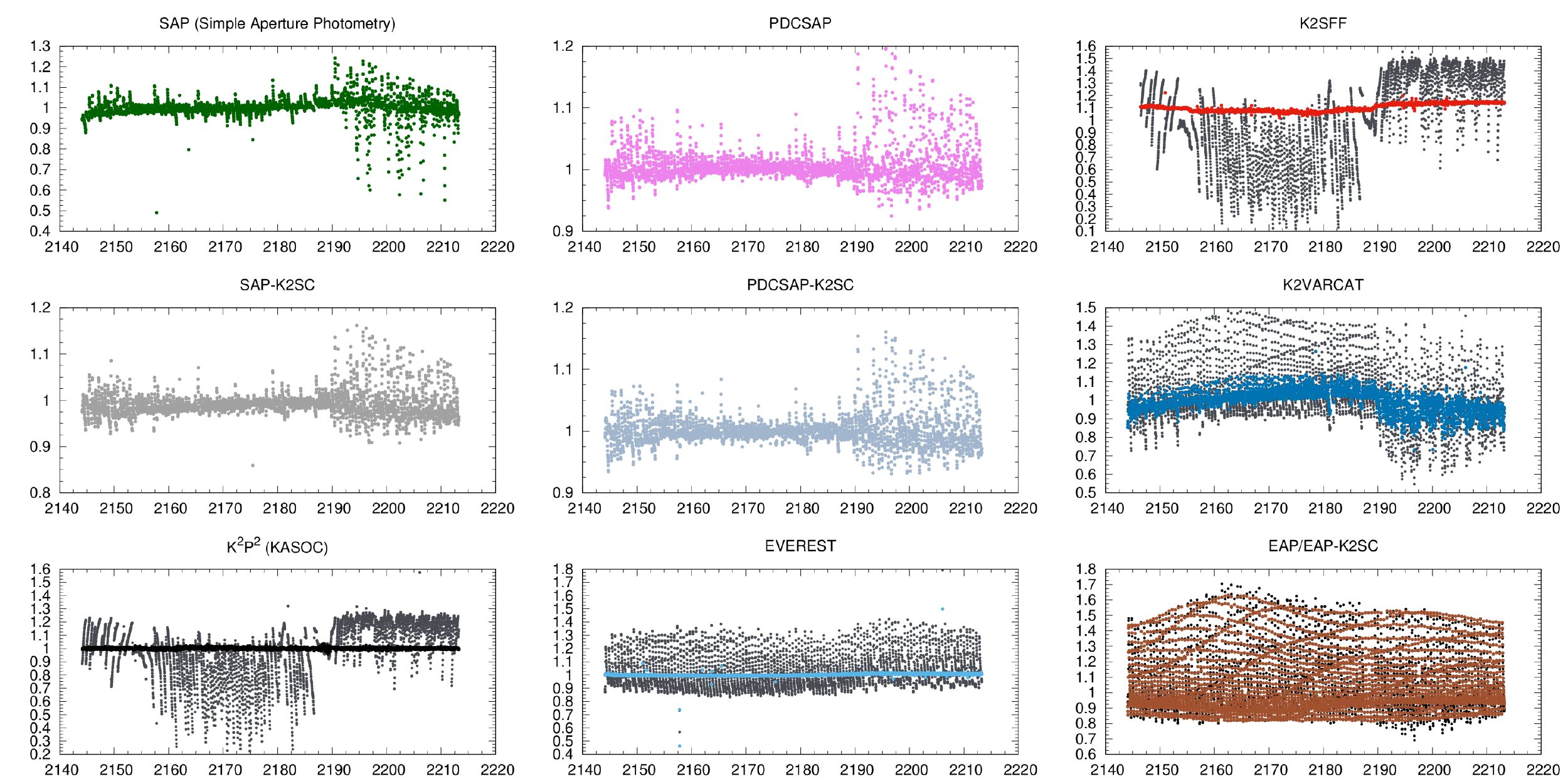}
\caption{Photometry of an RRab star (EPIC 206115430, \textit{Kp} = 16.512 mag) from various pipelines. 
Here the TPF contained the blended image of two stars, with the RR Lyrae target being the off-center one, and thus missed by multiple pipelines. (x-axis: BJD-2454833, y-axis: relative flux)} 
\label{fig:k2-example4}
\end{figure*}

We selected a few examples that show some of the problems various photometric pipelines encounter when dealing with RR~Lyrae light curves. Figures~\ref{fig:k2-example1}, \ref{fig:k2-example2}, \ref{fig:k2-example3} and \ref{fig:k2-example4} show four stars as measured by the official K2, and the K2SC-corrected SAP and PDCSAP photometry, the K2SFF, K2VARCAT, K2P$^2$, and EVEREST light curves, as well as our own EAP and K2SC-corrected EAP light curves. 

Bright stars (above about \textit{Kp} = 14--15 mag) are generally of good quality, at least concerning the raw photometry the various methods obtain. Our first example, EPIC 206144794 (\textit{Kp} = 12.667 mag, Campaign 3), in Fig.~\ref{fig:k2-example1}, has a strong beating between the pulsation period and the long-cadence sampling of \textit{Kepler}, and thus the light curve clearly shows if the photometry is degraded by errors. The SAP photometry indicates that the assigned aperture was suboptimal at the beginning and through the last 20 days of the campaign and the correction made by the PDCSAP algorithm is not sufficient. Further corrective measures by the K2SC pipeline get rid most of the instrumental noise, but some scatter still remains. The raw photometric data of the K2SFF, K2VARCAT and K2P$^2$ (grey points) all look acceptable, but the corrected light curves are severely degraded: in the case of K2SFF, even the presence of modulation becomes uncertain. The EVEREST pipeline, on the other hand, produces a smooth light curve very similar to the EAP one. 

We also selected two fainter stars to illustrate other instrumental issues found in other pipelines. EPIC~210438688 (\textit{Kp} = 17.387 mag) is a first-overtone star, observed during Campaign~4. It is apparent from the SAP light curve in Fig.~\ref{fig:k2-example2} that large amounts of flux spilled from the assigned aperture during the last third of the campaign. Closer inspection of the movements of the photocenter of the star revealed that when the direction of the attitude drift reversed, not only the amplitude of the pointing jitter got higher, but the mean position of the photocenter shifted as well, moving large parts of the PSF outside the aperture. This issue then propagated into the PDCSAP and K2SC light curves. The K2SFF, K2VARCAT, and K2P$^2$ light curves suffer from similar errors as in the previous case. The EVEREST pipeline failed to separate the instrumental and intrinsic signals and removed most of the pulsation signal. 

The third example is EPIC~210507908 (\textit{Kp} = 19.085 mag), a faint RRab star, also from Campaign~4 (Fig.~\ref{fig:k2-example3}). In this case the SAP data is barely recognizable. But more importantly, the average flux level of the SAP data is changing during the Campaign. Apparently, the PDCSAP algorithm assumes that this change is caused by excess flux from blending, and subtracts it from the SAP data (whereas flux loss from the aperture should be compensated by scaling). The resulting PDCSAP therefore shows pulsation amplitude changes over the campaign. However, comparison with (the raw photometry of) other pipelines and with the EAP light curve indicates that this amplitude change is actually spurious. Therefore the use of PDCSAP and its K2SC-corrected version both would lead to a false positive detection of the Blazhko effect.

A fourth star (EPIC 206115430, \textit{Kp} = 16.512 mag) from Campaign~3, is our example where the pipelines failed to find the target star in the mask (Fig.~\ref{fig:k2-example3}). The center of the mask falls to another, non-pulsating star with similar brightness that is blended with the RR Lyrae one at the resolution of \textit{Kepler}. The raw K2SFF and K2P$^2$ data are dominated with variations caused by photocenter motion, which was successfully corrected, but that led to flat light curves for the other source. The raw photometry of K2VARCAT and EVEREST was able to catch the RR Lyrae variability, but the modulation is hardly detectable in them. Their corrected versions failed here as well. The K2SC-corrected EAP light curve is clearly the best quality among them, and it revealed a complex modulation.

We constructed similar plots presented in Figures ~\ref{fig:k2-example1}, \ref{fig:k2-example2}, \ref{fig:k2-example3} and \ref{fig:k2-example4} for all targets, to estimate the success rate of our method. We found that the EAP-K2SC pipeline produced the best-quality light curves for nearly one third of the whole sample. The other EAP light curves are mostly equally good as the best light curves from the available pipelines, which is very often the raw EVEREST photometry or the K2SC-corrected PDCSAP. We found five cases where systematic variations were removed by EAP-K2SC the most efficiently, but the strong contamination was handled better by the other pipelines.

\subsection{Ensemble comparison between the pipelines}
While the examples above illustrate a number of problems with the other pipelines, they are not necessarily representative of the whole RR~Lyrae sample. Therefore we created two different benchmarks to measure the performance of the pipelines (strictly in relation to the RR Lyrae light curves:

\begin{enumerate}
\item{The median of the residual frequency spectrum between 1.0 and 4.0 d$^{-1}$ after prewhitening with the significant peaks (SNR $>$ 4). This is the frequency range where most pulsation signals occur, including the fundamental mode, the first overtone, and most low-amplitude additional modes. The red noise component of the instrumental signals also increases in this region. }
\item{The difference between the amplitude of the main frequency component in the EAP light curve and the other light curves (if present). This benchmark, $\Delta A_0 = A_{0(\rm other)}-A_{0(\rm EAP)}$, measures whether some of the pulsation amplitude has been lost during processing.}
\end{enumerate}

\begin{figure*}
\includegraphics[width=1\textwidth]{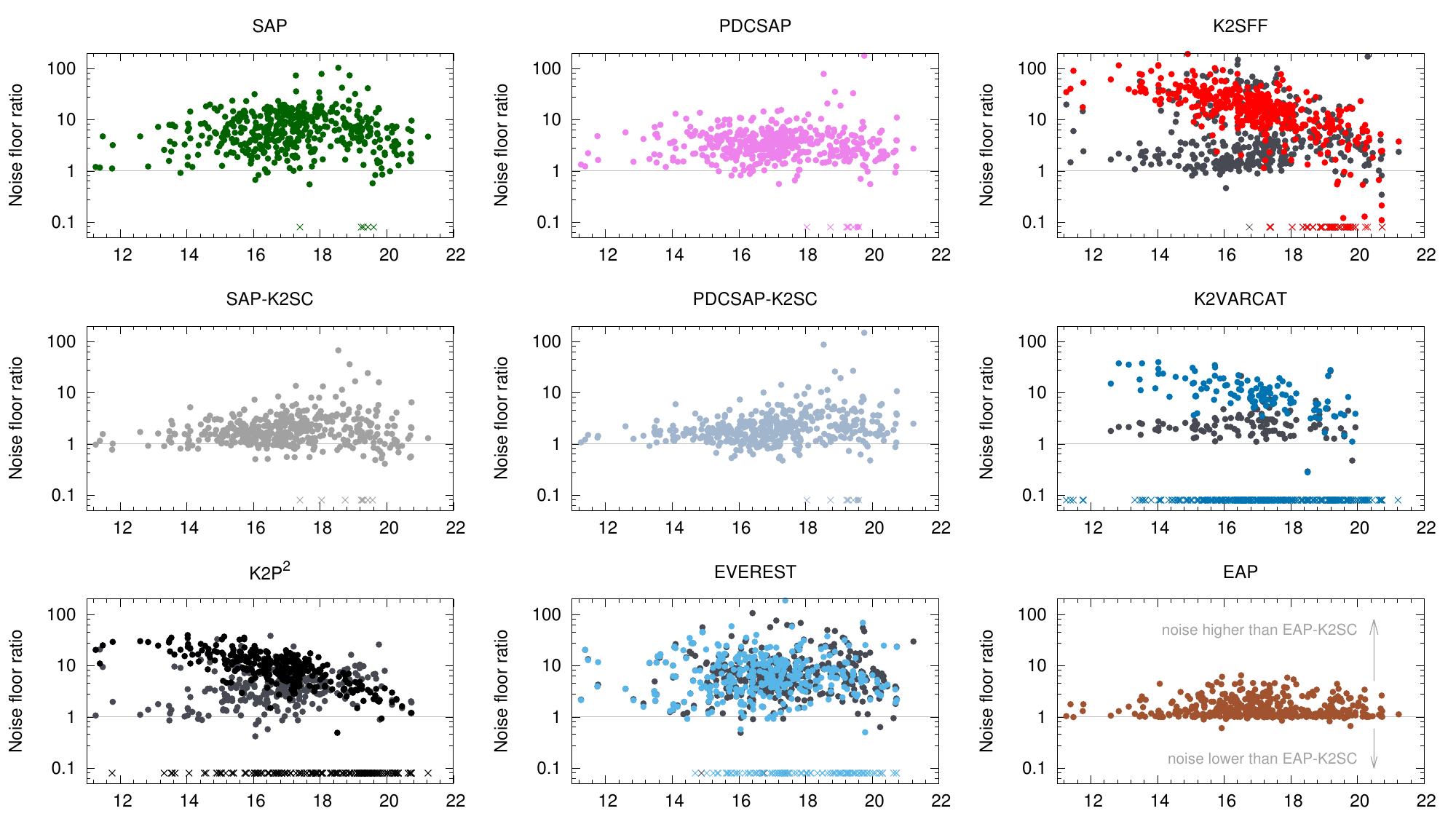}
\caption{Comparison of the noise floors, here defined as the median of the residual frequency spectra between 1.0 and 4.0 d$^{-1}$. The higher the value, the more pulsation signals may be lost compared to the EAP-K2SC photometry. Crosses mark stars missing from the respective databases. (x-axis: \textit{Kp} magnitude)}
\label{fig:eap-resspektr}
\end{figure*}

\paragraph{Noise floor: median of the residual spectra (Fig. \ref{fig:eap-resspektr})} The comparisons of the medians of the residual spectra shows that SSF-based methods increase the noise levels for RR Lyraes considerably in the low-frequency region. The difference is the highest for K2SFF and K2VARCAT, where the noise floors of the Fourier-spectra are elevated by the order of 2000 ppm, but may reach as much as 5000 ppm. For bright stars, that is almost a 100-fold increase compared to the noise floor of the EAP-K2SC data, and may easily hide otherwise obvious features, e.g., modulation, as in the case of EPIC 206144794 (Fig.~\ref{fig:k2-example1}). K2P$^2$ and EVEREST fare better, but they still increase the noise floor of the spectra roughly 10-fold for brighter RR Lyrae stars. The noise levels are only comparable at the faint end where photon noise starts to be the dominant source of photometric errors. 

The SAP and PDCSAP spectra show elevated noise for basically all brightness ranges, but the K2SC algorithm is able to compensate for that quite well down to about $Kp = 17$ mag. Here we also see a plurality of stars with noise floors below that of EAP-K2SC. This difference likely arises from the difference in aperture sizes, i.e., the smaller apertures used by the SAP and PDCSAP photometries gather less noise from pixels containing background flux than the larger EAP apertures. However, too small apertures may introduce other errors, as we have shown earlier. A comparison between the plain and K2SC-corrected EAP light curves indicate that while the improvements are noticeable, the differences are far smaller that those from the other pipelines. 

We also indicated the stars that are not present in the various pipelines with crosses at the bottom of each subplot. While the SAP/PDCSAP and related K2SC samples are almost complete, a large fraction of targets are missing from other pipelines, especially at the faint end. We note that at the time of this writing, K2VARCAT was only available up to Campaign 4.

\begin{figure*}
\includegraphics[width=1\textwidth]{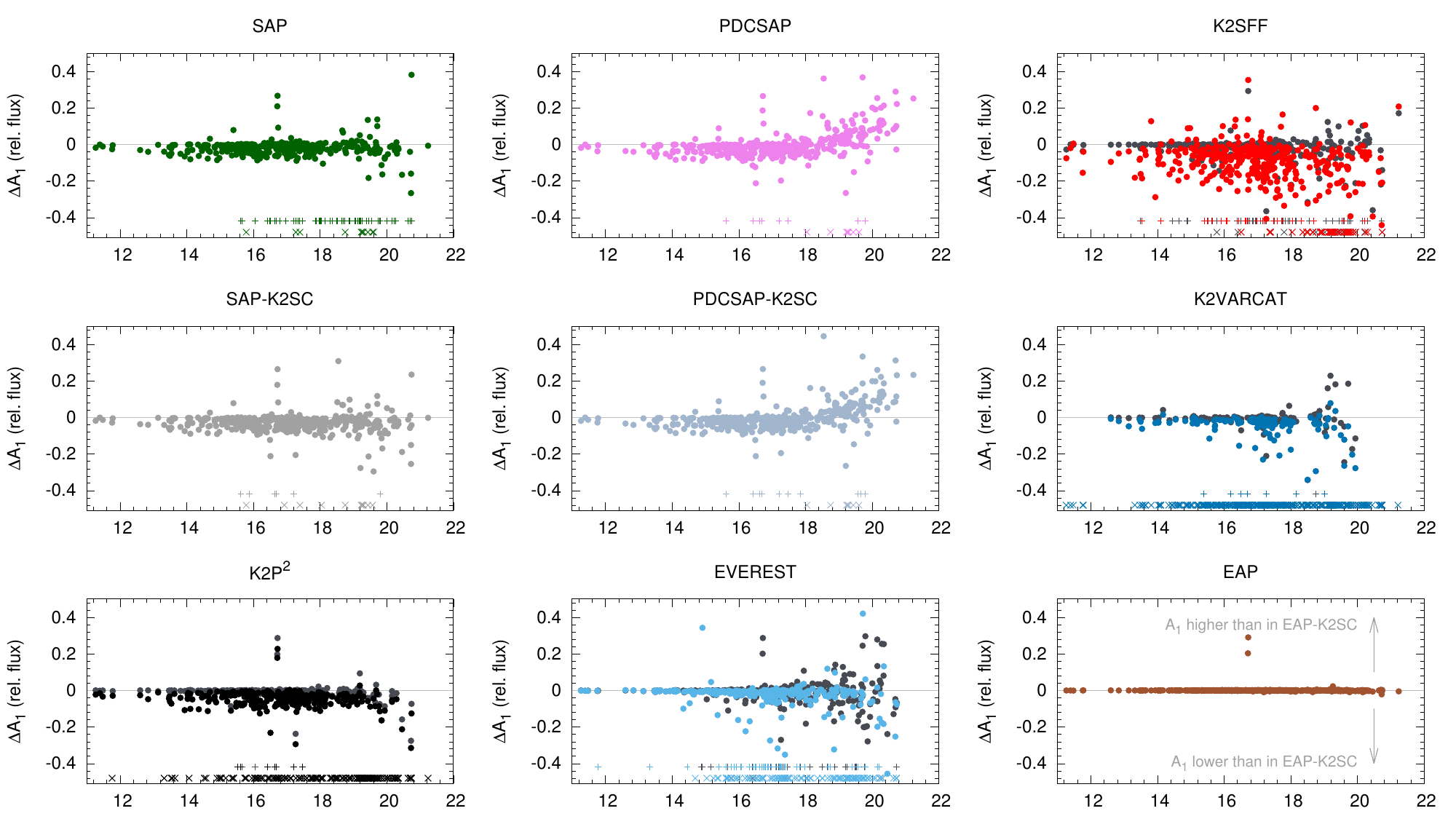}
\caption{Same as Fig.~\ref{fig:eap-resspektr} but for the differences in the amplitude of the fundamental mode frequency.  Crosses are missing data, plus signs indicate light curves where the pulsation frequency was not detected. (x-axis: \textit{Kp} magnitude) }
\label{fig:eap-mainpeaks}
\end{figure*}


\paragraph{Amplitude of the main peak (Fig.~\ref{fig:eap-mainpeaks})} The second measure we calculated is the difference of the relative flux amplitudes of the pulsation frequency peaks ($A_1$) determined from the EAP and the other light curves. Not surprisingly, coherent signals are strongly suppressed by the corrections used by K2SFF and K2VARCAT but are preserved relatively well by the other ones. Interestingly, PDCSAP, and its K2SC-corrected variant, show a clearly increasing excess towards the faintest targets.
Finally, here we indicated the stars where processed light curves exist, but the main peak could not be identified with plus signs. This issue, i.e., losing the pulsation signal, affects the SAP and EVEREST samples heavily. 
 
\begin{figure}
\includegraphics[width=0.48\textwidth]{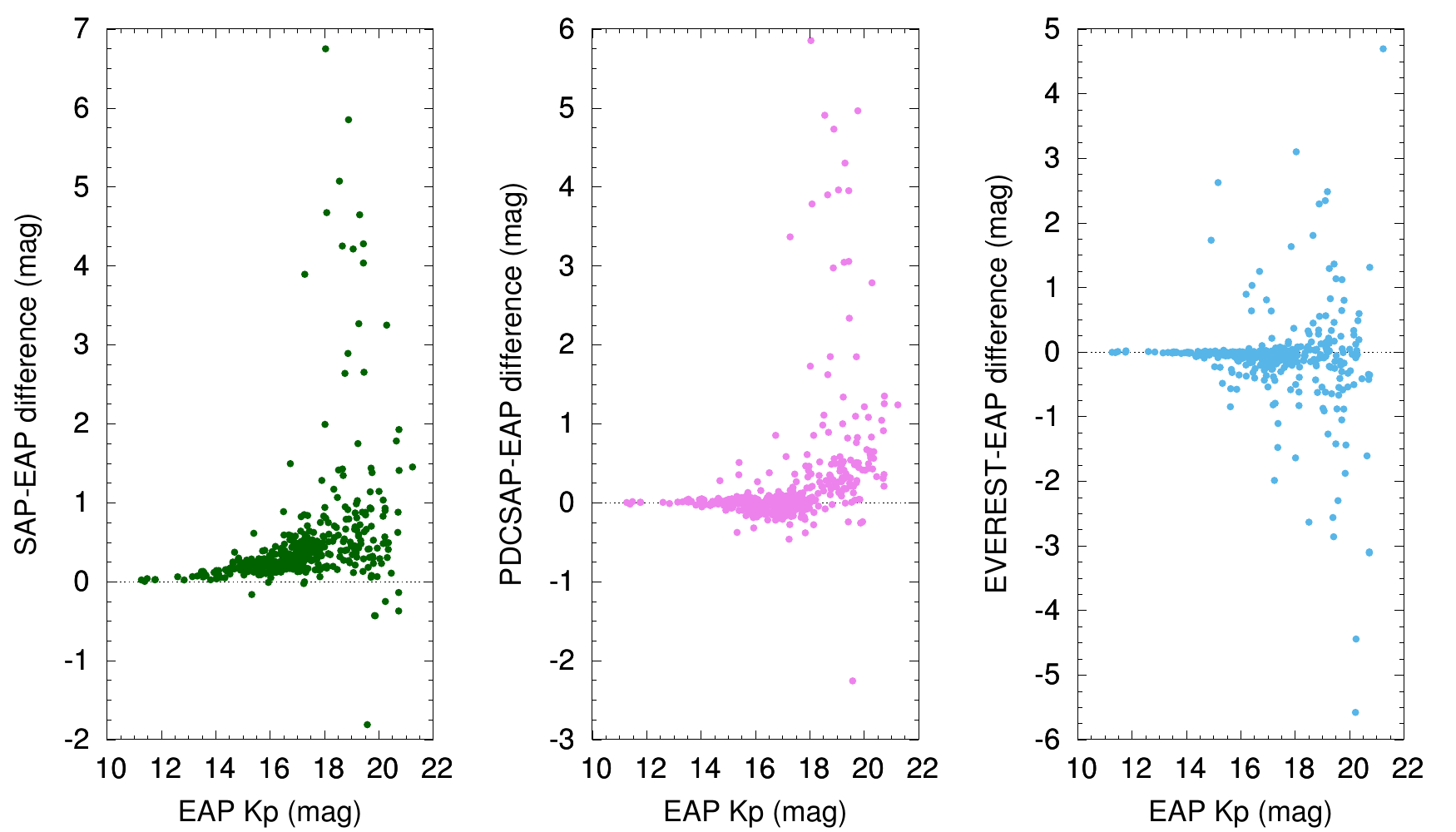}
\includegraphics[width=0.48\textwidth]{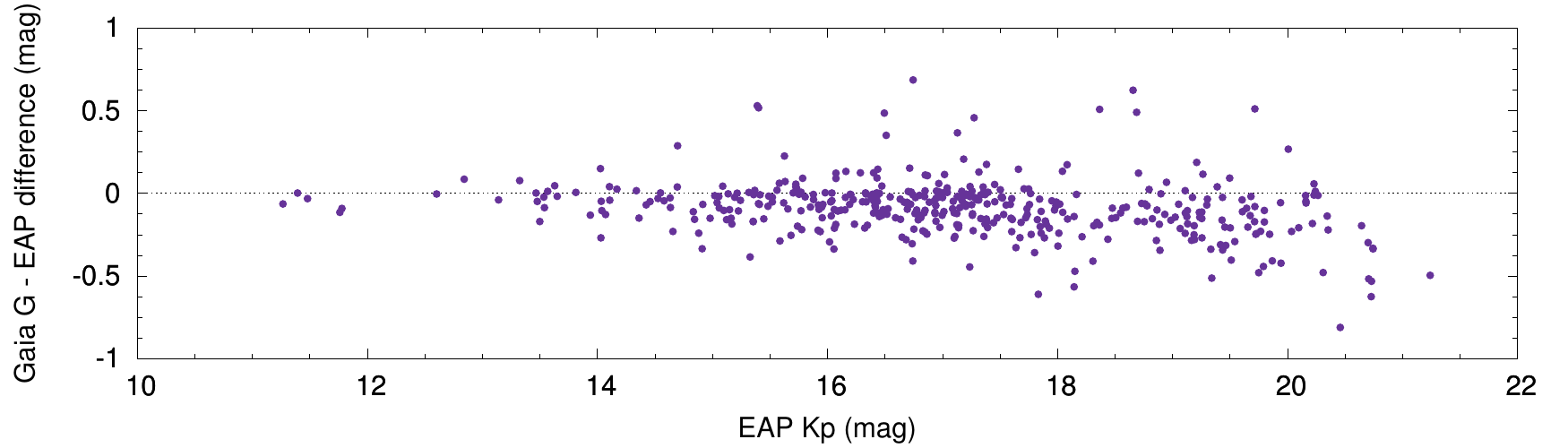}
\caption{Top row: Brightness differences from the EAP light curves, with the light curve median flux values converted to magnitudes. The SAP and PDCSAP light curves appear much fainter towards the faint end than the EAP values, whereas for EVEREST, there is scatter in both directions. Bottom: differences between the EAP and \textit{Gaia} \textit{G} values. }
\label{fig:magdiff}
\end{figure}

In order to understand the amplitude differences in Fig.~\ref{fig:eap-mainpeaks} we plotted the median flux brightness values of the SAP/PDCSAP and EVEREST light curves relative to EAP median brightnesses, all converted to \textit{Kp} magnitudes using the 25.3 mag zero point determined by \citet{k2p2-2}, in Fig.~\ref{fig:magdiff}. We emphasize that these values are based on the light curves and are different from the \textit{Kp} values listed in the EPIC catalog. It is clear that our values are systematically higher compared to the SAP values, and for stars fainter than \textit{Kp} = 18--19 mag we measure many of the targets several magnitudes brighter than both the SAP and PDCSAP pipelines. The origin of these large differences is not straightforward to determine: it could either mean that the SAP/PDCSAP light curves underestimate the average brightness of the star or that the large EAP apertures are contaminated by flux from the background pixels and/or other nearby sources. To estimate contamination, we crossmatched the sources with the \textit{Gaia} DR2 catalog and plotted the differences between the EAP \textit{Kp} and the \textit{Gaia} $G$ band mean brightness values (lower panel of Fig.~\ref{fig:magdiff}). As the $G$ and \textit{Kp} bands have very similar transmission curves, this comparison is not affected by color differences. The plot shows only half a magnitude difference on average between the \textit{Gaia} brightnesses and our reductions. Therefore we can conclude that the large differences originate from the SAP and PDCSAP pipelines. These use rather small apertures that only capture a portion of the image of the star, and thus only a fraction of the stellar flux. Moreover, as large parts of the PSF end up in the background pixels, they elevate the background level and reduce the median flux of the star, leading to larger perceived amplitudes. This gets more pronounced at the faint end where apertures are sometimes as small as a single pixel. Comparing the EAP brightnesses to the EVEREST ones, we notice a large scatter for stars fainter than $\sim$16 magnitude in both directions, but no systematic deviation can be reported.

Overall, these benchmarks show that the EAP pipeline provides substantially better photometry than the other methods for a large number of stars. 

\section{Analysis and results}

While the main purpose of this paper is to present the first batch of RR Lyrae EAP light curves observed in early K2 campaigns, we carried out some analysis of the database, mainly to inspect and classify the stars. 



\begin{figure*}
\includegraphics[width=1.0\textwidth]{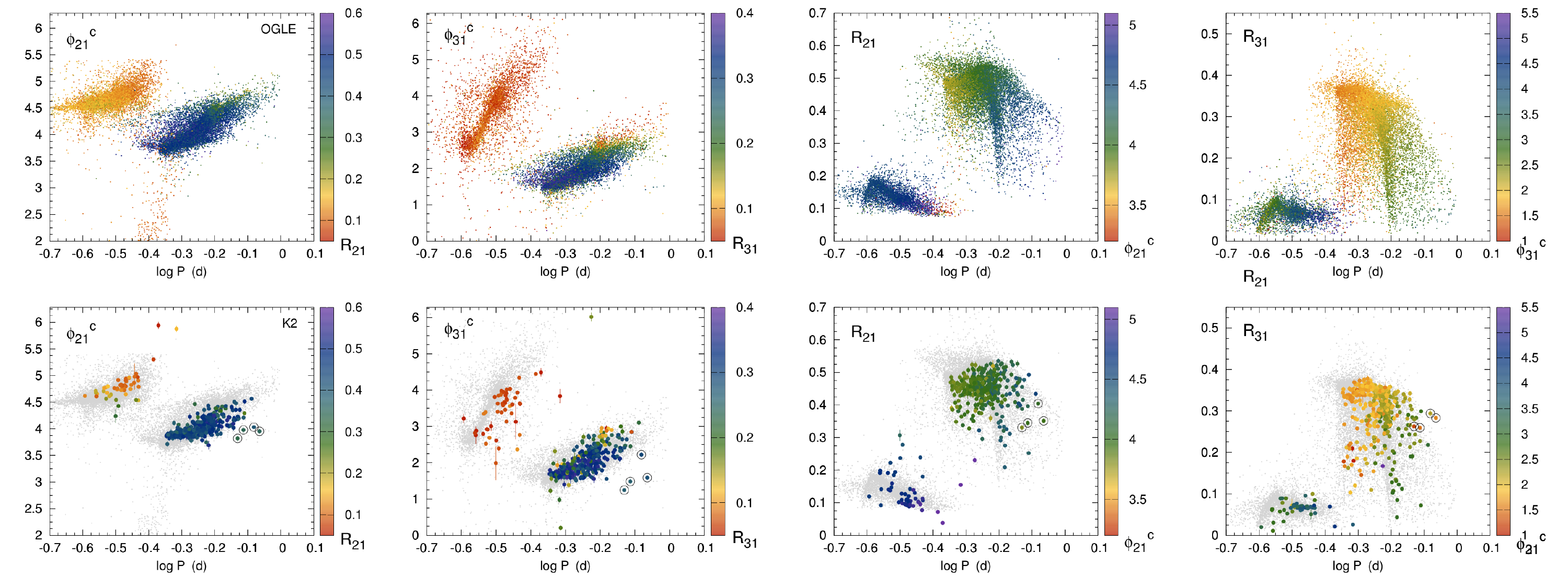}
\caption{Relative Fourier coefficients of the single-mode stars from the OGLE bulge sample, transformed to $V$ band (top), and the single-mode K2 RR Lyrae stars from this paper (bottom). Color coding shows the mutual relation between the $\phi_{i1}$ and R$_{i1}$ parameters. The OGLE sample is shown in light grey under the K2 sample. Anomalous Cepheid candidates are marked with black circles.  }
\label{fig:fourparam}
\end{figure*}

\subsection{Fourier parameters and classification}

It was shown by \citet{simon-teays} that low-order Fourier coefficients can describe the structural properties of RR Lyrae (and Cepheid) light curves efficiently. The four relative Fourier coefficients measure the amplitude ratios and phase differences between the first and second or third Fourier-terms, respectively: $R_{i1} = A_i/A_1$ and $\phi_{i1}=\phi_i-i\,\phi_1$, where $i$ is either 2 or 3. This technique is widely used for classification purposes, and to compare hydrodynamic models with observations, as it was done for the original \textit{Kepler} sample \citep{nemec2011}. Although most K2 light curves can be classified by simple visual inspection, there can be ambiguous cases, such as stars that fall between the RRab and RRc loci: one such example in the K2 data is V397~Gem \citep{molnar-ibvs}. Moreover, our sample can be contaminated by contact eclipsing binaries and high-amplitude $\delta$ Scuti stars at the short-period end, and by anomalous Cepheids and BL~Her stars towards the longest periods. Therefore we calculated the relative Fourier coefficients for our sample.

As for comparison, the best available source is the OGLE-IV catalog of the Bulge RR Lyrae stars \citep{ogleiv}. While most of the observations were made in the $I$ band, the coefficients can be converted into the $V$ coefficients with the equations derived by \citet{morgan1998}. The main drawback of this comparison is that the converted amplitude ratios contain artefacts for stars where the $R_{i1}$ values could not be determined from the $I$ band, as seen in the upper panels of Fig.~\ref{fig:fourparam}.

Although high-precision photometry revealed that many RR Lyrae stars are not single mode stars, here we refer to stars showing one large-amplitude, dominant mode as ``single-mode" stars. We compared the converted Fourier parameter values of OGLE RRab and RRc stars with the K2 stars in Fig.~\ref{fig:fourparam}. Most of the stars fall into the two main loci, in agreement with the loci of the OGLE stars.
We also identified non-pulsating variables and non-variable stars and listed them in Table~\ref{table:nonrrl} in the Appendix. The proper classification of these 25 stars is beyond the scope of this paper, but we indicated if we identified the signs of rotational variability or an eclipsing binary. There is one star that needs special attention, EPIC 121663528, which might be not a rotational variable but W Virginis-type star.

We found 371 ($\sim$85\%) RRab and 50 RRc ($\sim$12\%) stars in our sample, but we note that these ratios are likely biased towards RRab stars. Target selection for the Guest Observer proposals focused on the strong RR Lyrae candidates, and potential RRab targets are easier to recognize than RRc ones, of which a significant part turned out to be eclipsing binaries. We also found eleven classical double-mode (RRd) stars: these are subjects of detailed investigation, to be presented in a separate paper (Nemec at al. in prep.), together with the other RRd stars observed during the entire K2 mission.

We note that the [Fe/H] iron content index of RR Lyrae stars, a proxy for their metallicities, can also be determined from the $\phi_{31}$ coefficients and the pulsation periods, as it was done for the original \textit{Kepler} sample and the stars in the K2-E2 engineering run \citep{nemec2013,Molnar-2015b}. However, we decided to present the analysis of the more extended K2 sample in a standalone paper. 

\subsubsection{Anomalous Cepheid candidates}

\begin{figure*}
\includegraphics[width=1.0\textwidth]{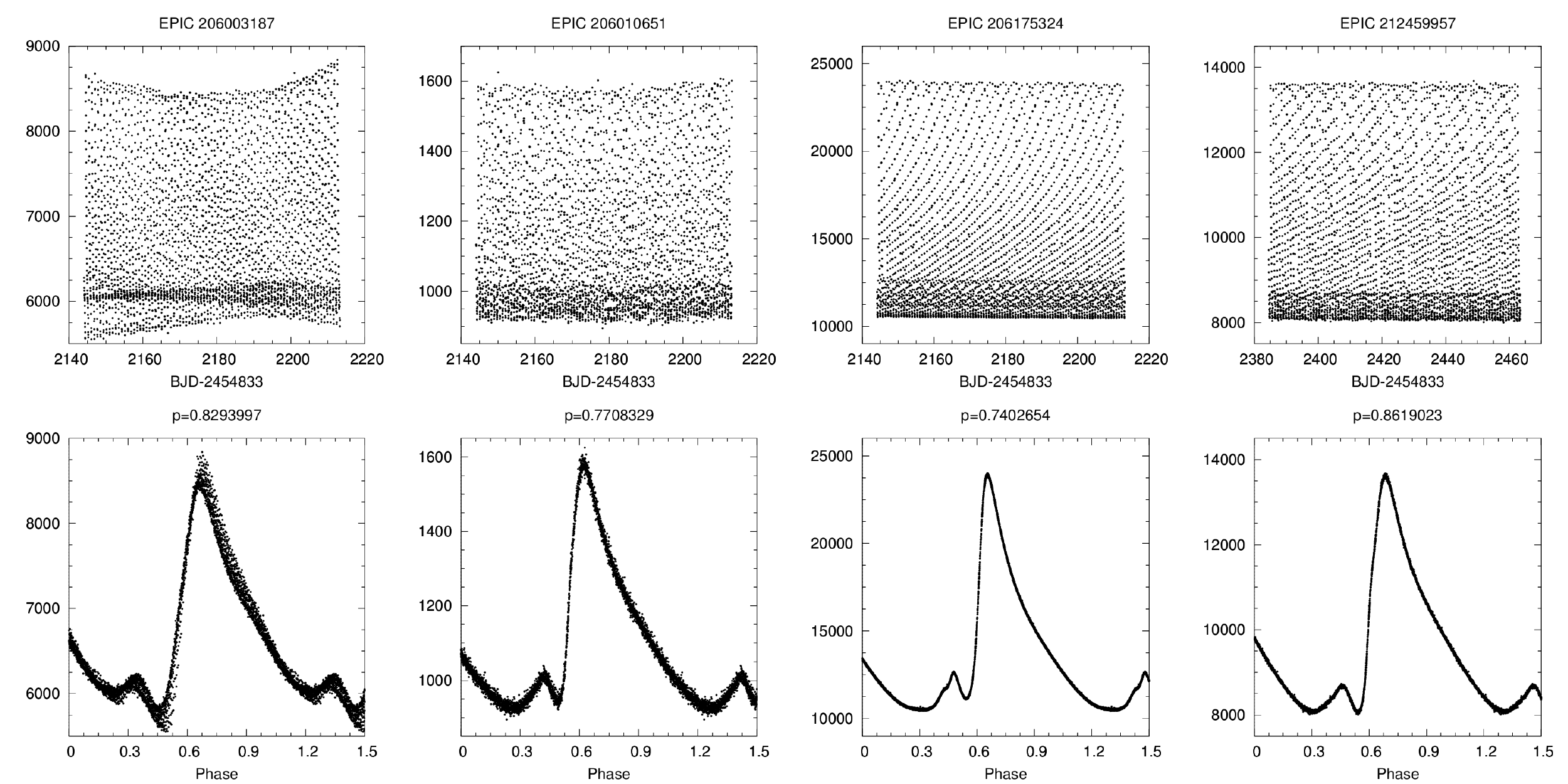}
\caption{EAP-K2SC light curves and phase curves of four anomalous Cepheid candidates. EPIC 206003187 shows a clear Blazhko-type modulation, while the extra variation in EPIC 206010651 is weak, and might be of instrumental origin. Values are in flux units (e$^-$/s).}
\label{fig:acep}
\end{figure*}

\begin{figure*}
\includegraphics[width=0.98\textwidth]{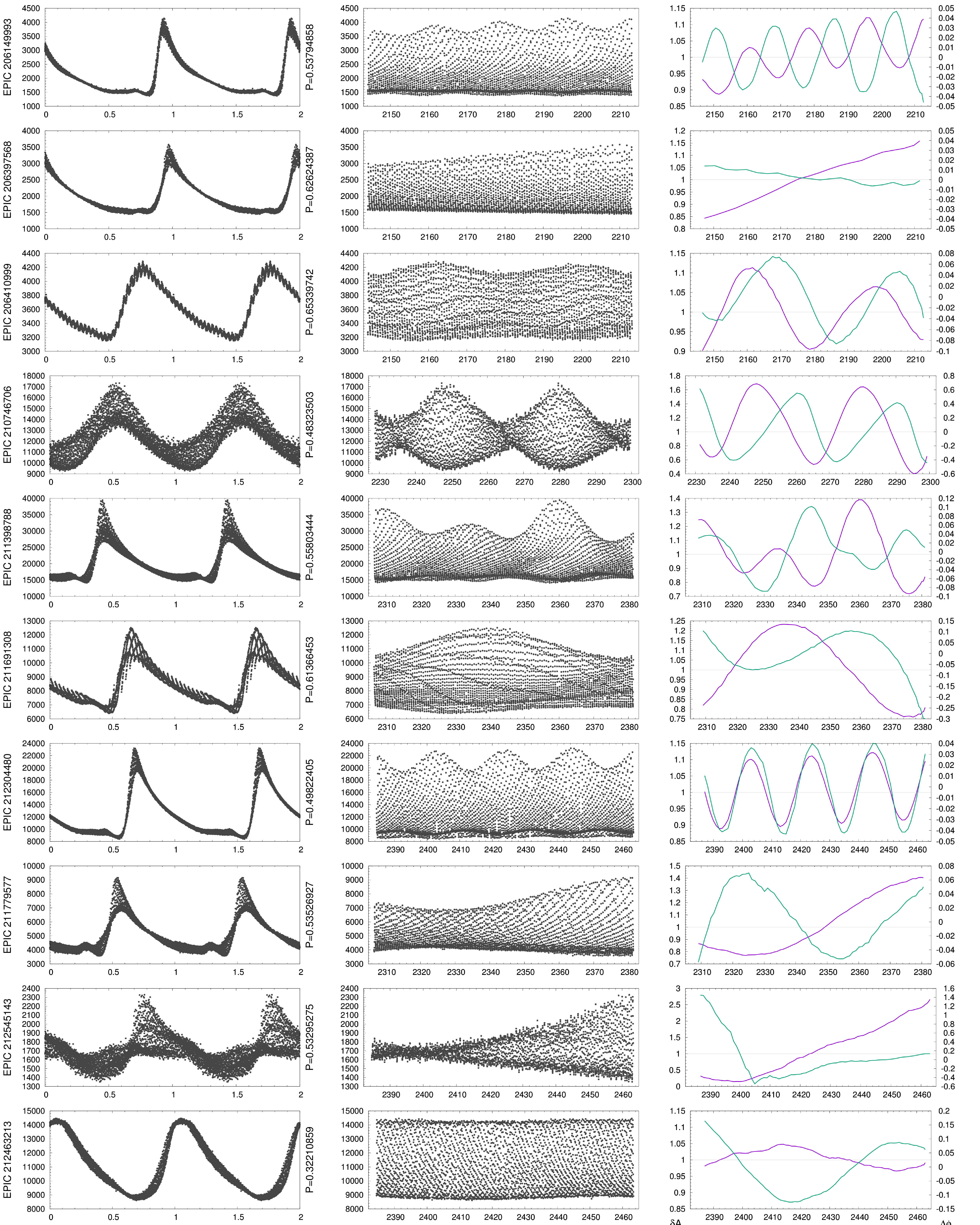}
\caption{Examples of modulated RR Lyrae stars. Left: phased light curves. Middle: EAP-K2SC light curves in flux (e$^-$/s). Right: amplitude modulation curve (purple), phase modulation curve (green). $\delta A$ is the amplitude scaling parameter, $\Delta\phi$ is the phase shift parameter of the template.} 
\label{fig:examples}
\end{figure*}

Anomalous Cepheids are a rare subtype of the family of classical pulsators. They occupy the parameter space between RR Lyrae and Cepheid stars, both in the $\phi_{i1}$, $R_{i1}$ planes in the period-luminosity relation \citep{mc-acep,blg-acep}. It is not always straightforward to separate them from RR Lyrae stars in the short-period range ($<1$ d), if no parallax information is available. We identified four strong candidates that we marked with black circles in Fig.~\ref{fig:fourparam}. Unfortunately, these stars are too faint to have useful parallax information in \textit{Gaia} DR2, therefore we cannot confirm that they follow the anomalous Cepheid PL-relation. 

Their light and phase curves presented in Fig.~\ref{fig:acep} have different shapes compared to that of RRab stars in the same period range, in accordance with their position in the Fourier parameter space, outside the RRab regime. Unfortunately, no-high precision Fourier parameter measurements are available for anomalous Cepheids, and no transformation is available yet between $I$-band and $V$-band Fourier parameters for them either in the OGLE Survey. Fortunately, a $V$-band study was produced by \citet{jur}, who used OGLE $V$-band data along with photometry from other large sky surveys for a hundred galactic short-period anomalous and type II Cepheids. Our four K2 anomalous Cepheid candidates fit into the fundamental-mode anomalous Cepheid (ACEP-F) group presented in that paper.

A remarkably clear Blazhko-type modulation is visible in EPIC 206003187, even though the cycle length appears to be considerably longer than the observation. If this star is indeed an ACEP-F, this is the first modulated one ever discovered. We also detect some extra variation in EPIC 206010651, but it is rather weak, and might have instrumental origin.

\subsection{Occurrence of amplitude and phase changes}

The incident ratio of Blazhko effect among RRab stars is believed to be $\sim$40-50\%. Recent studies of RR Lyrae stars in the Bulge measured by OGLE-IV  and in the original \textit{Kepler} field found it to be 40.3\% (of 8282 targets) and 51\% (of 35 stars) respectively \citep{prudil,Benko-2019}. This is in agreement with the landmark results of the Konkoly Blazkho Survey that found it to be 47\% (of 30 stars) and 43\% (of 105 stars) for short- and long-period RRab stars \citep{kbs,kbs2}. In contrast to these findings, \citet{kovacs2018} found the signs of modulation, in the form of frequency side peaks, in 91\% of 151 RR Lyrae stars in the C1-C4 Campaigns of K2 mission. A subset of the targets overlap with our sample in C3--C4.

Beside the visual inspection of the light curves we used two more rigorous approaches to recover modulation: we searched for modulation triplets in the Fourier spectra, and we constructed amplitude and phase variation curves with a template fitting method. Our templates were the five-order harmonic series for each light curve, where the Fourier amplitudes and phases were scaled and shifted with a single parameter, with $A_i = \delta A\cdot A_{i0}$ and $\phi_i = \phi_{i0} - \Delta \phi$, respectively. Some examples of these modulation parameters, along with the respective light curves, are shown in Fig.~\ref{fig:examples}. This latter approach provides the opportunity to estimate the periods of the amplitude and the phase modulation independently, by fitting sine functions to the two modulation curves separately. In several cases we observe only non-repeating or monotonic changes in the amplitude and phase over the length of a Campaign. Since the only known case for smooth, large-scale changes in the pulsation cycles of RR Lyrae stars is the Blazhko effect, we included these stars into the Blazhko-modulated group as well.  

We report our findings, along with the numbers of the different subtypes, in Table \ref{stat}. Note that we included the four ACEP candidates among the RRab stars as their classification is still uncertain.

\begin{table}[h!]
\renewcommand{\thetable}{\arabic{table}}
\centering
\caption{Number of targets by type.} \label{tab:statistics}
\begin{tabular}{lccc|c}
\label{stat}
\tablewidth{0pt}

Campaign & RRab(Bl)&  RRc(Bl)& RRd & Total\\  
\hline
C3 &78(41) & 0& 0 & 78\\
C4 &66(22) & 5& 4 & 75\\
C5 &67(30)& 15& 3 & 85\\
C6 &160(73) & 30(2)& 4& 194\\
\hline
Sum&371(166)  &50(2) & 11 &432\\
\end{tabular}

\end{table}

\begin{figure}
\includegraphics[width=0.48\textwidth]{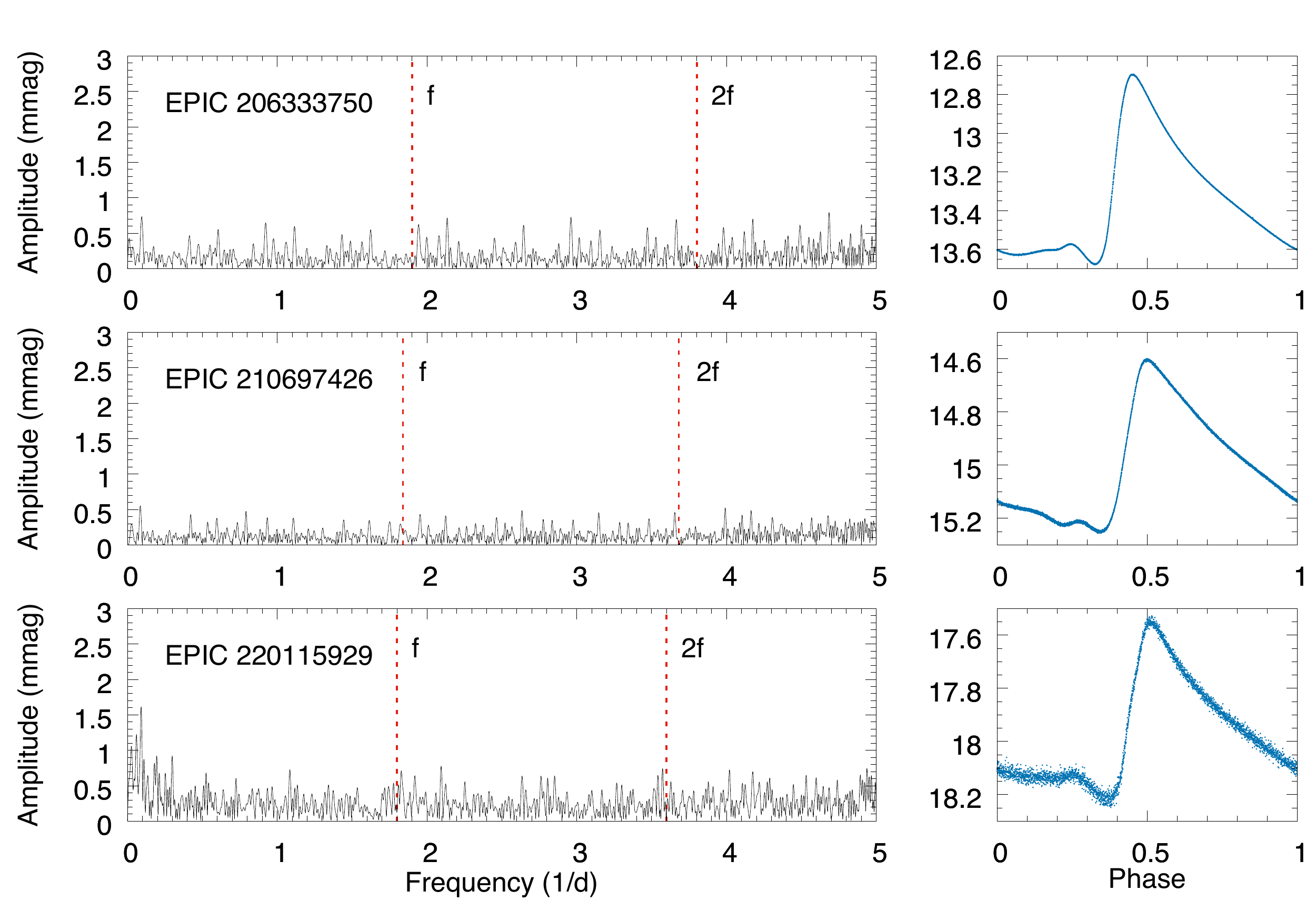}
\caption{The lack of modulation in non-Blazhko stars. Left: the residual Fourier spectra after prewithening with the main pulsation frequency and its harmonics. Right: folded light curves plotted in \textit{Kp} mag. The instrumental noise and trends are more perceptible in the faint one.} 
\label{fig:nonbl}
\end{figure}

\begin{figure*}
\includegraphics[width=0.98\textwidth]{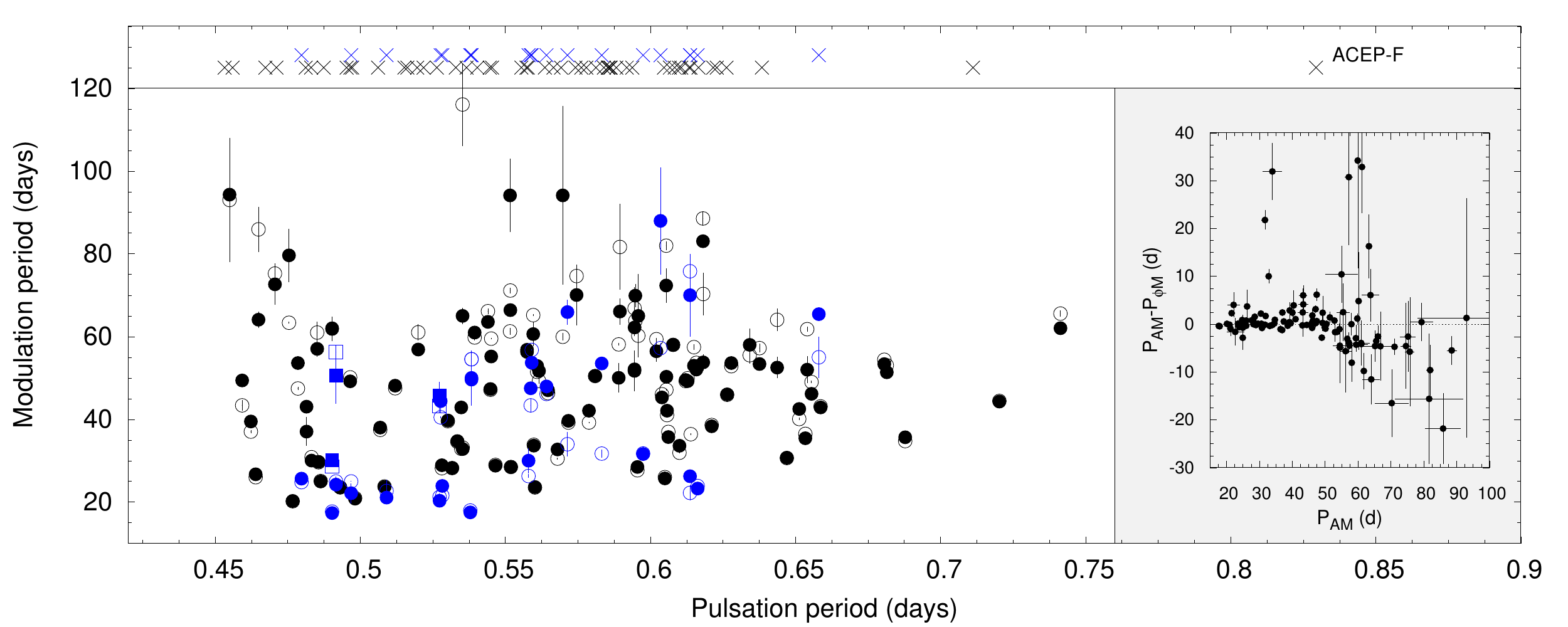}
\caption{Periods of amplitude modulation (open symbols) and phase modulation (filled symbols) against the pulsation periods. Blue symbols denote stars with multiple modulation, with squares showing the secondary modulation periods where they could be determined. Crosses mark the stars where the Blazhko period is too long to estimate. The modulated ACEP-F candidate is in the upper right corner. The insert shows the differences between the amplitude and phase modulation periods against the amplitude modulation period.}
\label{fig:bl}
\end{figure*}

We found 44.7\% of the RRab stars (166/371) to be either clearly modulated or show clear amplitude and phase changes: they show both the modulation triplets in the Fourier spectra and variation in the amplitude and phase curves. We note that the phase modulation is very weak in some cases, compared to the amplitude changes. We found two RRc stars with amplitude and phase changes as well, giving a ratio of 4\% among the overtone stars. One of the Blazhko RRc stars is presented in the bottom row of Fig.~\ref{fig:examples}.

We also found 18 stars that have weak additional frequencies at one side of the main pulsation peak. We cannot distinguish if those peaks correspond to small intrinsic changes or are the results of contamination, therefore we did not mark these uncertain identifications as Blazhko stars. However, even if they were all \textit{bona fide} modulated stars, the ratio would still not exceed 50\%. The remaining 187 stars show clean residual spectra after removing the Fourier series of the main pulsation. We present three examples of these non-Blazhko stars at different brightness in Fig.~\ref{fig:nonbl}.

The periods determined for the amplitude and phase modulations, respectively, are presented in Fig.~\ref{fig:bl}, plotted as the function of the pulsation period. No correlation is apparent between the pulsation and modulation periods, except for a slight trend in the lower limit of the distribution, with the shortest modulation periods occurring towards the shorter pulsation periods. This is in agreement with the findings of \citet{skarka2016}.

We also identified several examples of variable or multiperiodic modulation, but unfortunately we were not able to estimate modulation periods longer than the observation length, except for some clear cases when sufficient extrema were available.  We also compared the periods fitted to the amplitude and phase changes for each star individually. We concluded that below approximately 55 days there is little discrepancy between the two values, except for a few outliers. Above that, the results are more uncertain, for two reasons. First, the periods are nearing the length of the data, and second, the shapes of the two modulations start to differ more prominently. Two examples for this issue are EPIC 211691308 and 211779577 in Fig.~\ref{fig:examples}. The phase modulation curves, in particular, become rather diverse in shape, compared to the amplitude changes, and thus their periods also become more uncertain. We show the progression of the period difference against the amplitude modulation period in the insert of Fig.~\ref{fig:bl}. 

Although the lengths of the Campaigns are too short to draw further conclusions about the correlation between the amplitude and the phase modulations, it is clear  that they are not always strictly correlated, especially towards longer Blazhko periods. Changes in the relation between the amplitude and phase modulations were reported for a few stars before \citep{v445lyr,nemec2013}. Based on those, we can expect these variations to average out over time, lessening the large apparent differences in the Blazhko periods we observe here.

\section{Summary and conclusions}
Developing photometric pipelines that properly correct for instrumental effects but is also able to preserve a wide variety of stellar variations is undoubtedly a huge challenge. RR~Lyrae stars, in particular, can be notoriously hard to handle, given their large pulsation amplitudes, short periods, and distinctive, sawtooth-shaped light curves. In this paper we provided a short, not exhaustive review of the various general-purpose algorithms and pipelines developed by various groups to correct the instrumental effects present in the K2 mission of the \textit{Kepler} space telescope. Most methods provide good raw photometry for these stars, often exceeding the quality of that of the official SAP/PDCSAP light curves. However, since the pulsations happen on time scales similar to the attitude changes, the SFF-based algorithms (K2SFF, K2VARCAT, K2P$^2$) usually fail to separate the two variations and provide bogus results for these kinds of stars. Methods based on Gaussian processes instead of simple decorrelation fare better: however, even the EVEREST pipeline only manages to correct about half of the RR Lyrae stars properly. Finally, we found that the only pipeline that can reliably correct RR~Lyrae light curves is K2SC, the other Gaussian process-based method, but unfortunately it relies on the SAP/PDCSAP light curves as input.

Instead of relying on these light curves, we decided to develop our own photometric solution called Extended Aperture Photometry (EAP), specifically tailored to RR Lyrae stars. We manually determined new pixel apertures for the targeted RR Lyrae stars in Campaigns 3--6, and then applied the K2SC correction to these new light curves \citep{k2sc}. Our tests show that the K2SC-corrected EAP light curves generally produce as good or better light curves for RR Lyrae stars as the best-performing other methods, but in a consistent way. 

We then used these light curves to classify the stars into their respective RR Lyrae subgroups (or otherwise determine if they are other types of stars), using the OGLE relative Fourier parameter values as comparison. We identified 432 RR Lyrae stars in the four Campaigns, but this is not an exhaustive list. Some further stars may hide as potentially misclassified eclipsing binaries or rotational variables in other proposals, or as background objects in the TPFs of unrelated targets. 

In this paper we focused on the photometry and did not carry out a detailed analysis of the stars, except for determining the presence and occurrence rate of the Blazhko effect. We found the rate to be below 50\%, contrary to the recent result of \citet{kovacs2018}, and in agreement with other studies \citep{kbs,prudil,Benko-2019}. We estimated the periods of the phase and amplitude modulations and found some differences above $\sim$55 days modulation period, but we are limited by the length of the Campaigns. We did not study additional low amplitude modes, but tested whether they can be recovered from our photometry efficiently. 

Both the raw EAP and the K2SC-corrected light curves are available for further analysis\footnote{\url{https://konkoly.hu/KIK/data_en.html}}. We plan to expand our method to the full K2 sample. Work is in progress to automate the method in order to process the thousands of RR Lyrae stars observed in later campaigns.  

\facilities{\textit{Kepler}/K2 \citep{howell2014}}

\software{PyRAF, PyKE \citep{pyke,pyke3}, K2SC \citep{k2sc2,k2sc}, lcfit \citep{lcfit}, Period04 \citep{period04}, gnuplot, numpy \citep{numpy}, lightkurve \citep{lightkurve}}

\section*{Acknowledgements}

The research leading to these results have been supported by the the Hungarian National Research, Development and Innovation Office (NKFIH) grants K-115709 and PD-121203, and the Lend\"ulet LP2014-17 and LP2018-7/2018 grants of the Hungarian Academy of Sciences. E.P.\ was supported by the J\'anos Bolyai Research Scholarship of the Hungarian Academy of Sciences. L.M.\ was supported by the Premium Postdoctoral Research Program of the Hungarian Academy of Sciences. M.S.\ acknowledges the Postdoc@MUNI project CZ.02.2.69/0.0/0.0/16--027/0008360. This work was performed in part under contract with the Jet Propulsion Laboratory (JPL) funded by NASA through the Sagan Fellowship Program executed by the NASA Exoplanet Science Institute. This paper includes data collected by the K2 mission, proposed by the RR Lyrae and Cepheid Working Group of the \textit{Kepler} Asteroseismic Science Consortium. Funding for the \textit{Kepler} and K2 missions are provided by the NASA Science Mission Directorate. This work made use of PyKE, a software package for the reduction and analysis of Kepler data, developed and distributed by the NASA Kepler Guest Observer Office. PyRAF is a product of the Space Telescope Science Institute, which is operated by AURA for NASA. The data presented in this paper were obtained from the Mikulski Archive for Space Telescopes (MAST). STScI is operated by the Association of Universities for Research in Astronomy, Inc., under NASA contract NAS5-26555. Support for MAST for non-HST data is provided by the NASA Office of Space Science via grant NNX09AF08G and by other grants and contracts. This research has made use of the KASOC database, operated from the Stellar Astrophysics Centre (SAC) at Aarhus University, Denmark. Funding for the Stellar Astrophysics Centre (SAC) is provided by The Danish National Research Foundation. This work has made use of data from the European Space Agency (ESA) mission \textit{Gaia} (\url{https://www.cosmos.esa.int/gaia}), processed by the \textit{Gaia} Data Processing and Analysis Consortium (DPAC, \url{https://www.cosmos.esa.int/web/gaia/dpac/consortium}). Funding for the DPAC has been provided by national institutions, in particular the institutions participating in the {\it \textit{Gaia}} Multilateral Agreement. We thank Suzanne Aigrain for initial discussions about the K2SC pipeline. The authors gratefully acknowledge the \textit{Kepler} Guest Observer Office for their tireless efforts which have made these results possible.




\appendix

Here we provide the Tables listing the RR Lyrae stars identified in Campaigns 3--6 (Table \ref{tab:maintable}), and those that turned out to be other types of stars (Table \ref{table:nonrrl}).

\begin{table*}[!h]
\caption{Table of RR Lyrae stars processed with the EAP method. The full table is available in the online version of the paper.}
\begin{center}
\label{tab:maintable}
\renewcommand{\arraystretch}{1.2}
\begin{tabular}{cccccccc}
\hline
\hline
EPIC & $\alpha_{2000}$ (deg) & $\delta_{2000}$ (deg) & Kp mag & Campaign & Type & FM period & O1 period \\
\hline
205905693	&	338.887895	&	-17.971128	&	14.362	&	C3	&	RRAB-BL	&	0.486393 &--			\\
205914832	&	337.438790	&	-17.615179	&	17.446	&	C3	&	RRAB-BL	&	0.528295 &--			\\
205915657	&	338.157238	&	-17.585205	&	17.143	&	C3	&	RRAB-BL	&	0.607875 &--			\\
205930577	&	337.107618	&	-17.032530	&	15.138	&	C3	&	RRAB	&	0.731083 &--			\\
205941274	&	332.566231	&	-16.662995	&	13.501	&	C3	&	RRAB	&	0.550847 &--			\\
205943711	&	333.047171	&	-16.579656	&	16.205	&	C3	&	RRAB	&	0.696769 &--			\\
205955110	&	338.043497	&	-16.204663	&	15.164	&	C3	&	RRAB-BL	&	0.594567 &--			\\
205969073	&	340.659483	&	-15.758557	&	13.540	&	C3	&	RRAB	&	0.509065 &--			\\
205977366	&	334.017926	&	-15.500817	&	17.604	&	C3	&	RRAB	&	0.524658	 &--		\\
205985225	&	341.824658	&	-15.263447	&	16.831	&	C3	&	RRAB-BL	&	0.654061 &--			\\
... & & & & & & \\
\hline

\end{tabular}
\end{center}
\end{table*}

\begin{table*}
\caption{Non RR Lyrae stars in the sample.}
\begin{center}
\renewcommand{\arraystretch}{1.2}
\label{table:nonrrl}
\begin{tabular}{cccccc}
\hline	
\hline	
EPIC & $\alpha_{2000}$ (deg) & $\delta_{2000}$ (deg) & Kp mag & Campaign & Type   \\

\hline
206090129	&	337.984710	&	-12.355544	&	13.702	&	C3	&	ROT	\\
206207481	&	332.688375	&	-9.637619	&	10.659	&	C3	&	EB	\\
212235345	&	338.731900	&	-13.070770	&	17.1	&	C3	&	ROT	\\
\hline											
210361885	&	62.215635	&	+12.541252	&	16.588	&	C4	&	EB	\\
210739713	&	66.577800	&	+18.953010	&	15.757	&	C4	&	EB	\\
210796097	&	58.836870	&	+19.809991	&	15.58	&	C4	&	EB	\\
210810957	&	62.425305	&	+20.034501	&	15.142	&	C4	&	EB	\\
210889448	&	62.124195	&	+21.226526	&	15.422	&	C4	&	EB	\\
210896547	&	53.537625	&	+21.338052	&	16.512	&	C4	&	ROT	\\
211101195	&	60.880065	&	+24.580213	&	15.418	&	C4	&	ROT	\\
\hline
211310837	&	133.971120	&	+10.069017	&	13.225	&	C5	&	EB	\\
211700735	&	132.465810	&	+15.925779	&	16.204	&	C5	&	?	\\
228682503	&	133.003820	&	+16.968360	&	20.01	&	C5	&	?	\\
\hline											
212308231	&	207.833955	&	-17.285618	&	15.642	&	C6	&	EB	\\
212321981	&	206.651835	&	-16.889869	&	15.119	&	C6	&	EB	\\
212421319	&	203.446320	&	-14.433414	&	16.407	&	C6	&	EB	\\
212502064	&	201.460275	&	-12.690578	&	9.671	&	C6	&	EB	\\
212504059	&	199.237995	&	-12.648150	&	11.601	&	C6	&	EB	\\
212522016	&	206.879760	&	-12.266881	&	17.498	&	C6	&	?	\\
212532428	&	202.193520	&	-12.043144	&	16.279	&	C6	&	?	\\
212634341	&	208.864890	&	-9.806615	&	13.552	&	C6	&	?	\\
212652164	&	208.934610	&	-9.396076	&	17.92	&	C6	&	?	\\
212663528	&	203.945760	&	-9.131430	&	15.278	&	C6	&	ROT/W Vir\\
212833339	&	204.089595	&	-4.454131	&	13.8	&	C6	&	?	\\
229228122	&	209.391660	&	-13.741389	&	19.17	&	C6	&	?	\\
\hline	
\end{tabular}
\tablecomments{Kp values are from EPIC.}
\end{center}
\end{table*}

\end{document}